\documentclass[10pt,preprint]{emulateapj}

\usepackage{graphicx}
\usepackage{lipsum}
\usepackage{gensymb}
\usepackage{placeins}
\usepackage{float}

\shorttitle{The Geometry of the \ion{Mg}{2}-absorbing CGM at $z\sim1.2$}
\shortauthors{Lundgren, Creech et al.}

\begin{document}

\title{The Geometry of Cold, Metal-Enriched Gas around Galaxies at $z\sim1.2$}

\author{Britt F. Lundgren\altaffilmark{1},
Samantha Creech\altaffilmark{1}, Gabriel Brammer\altaffilmark{2,3}, Nathan Kirse\altaffilmark{1}, Matthew Peek\altaffilmark{1,4},  David Wake\altaffilmark{1}, Donald G. York\altaffilmark{5}, John Chisholm\altaffilmark{6,7}, Dawn K. Erb\altaffilmark{8}, Varsha P. Kulkarni\altaffilmark{9}, Lorrie Straka\altaffilmark{10}, Christy Tremonti\altaffilmark{11}, Pieter van Dokkum\altaffilmark{12}}

\altaffiltext{1}{Department of Physics and Astronomy, University of North Carolina Asheville, Asheville, NC 28804, USA}
\altaffiltext{2}{Cosmic Dawn Center (DAWN)}
\altaffiltext{3}{Niels Bohr Institute, University of Copenhagen, Lyngbyvej 2, DK-2100 Copenhagen, Denmark}
\altaffiltext{4}{Space Telescope Science Institute, 3700 San Martin Drive, Baltimore, MD 21218, USA}
\altaffiltext{5}{Department of Astronomy and Astrophysics, University of Chicago, Chicago, IL 60637, USA}
\altaffiltext{6}{Hubble Fellow}
\altaffiltext{7}{Department of Astronomy, University of Texas at Austin, Austin Texas 78712-1205}
\altaffiltext{8}{Department of Physics, University of Wisconsin-Milwaukee, P.O. Box 413, Milwaukee, WI 53201, USA}
\altaffiltext{9}{Department of Physics and Astronomy, University of South Carolina, Columbia, SC 29208, USA}
\altaffiltext{10}{Leiden Observatory, Leiden University, PO Box 9513, 2300 RA Leiden, The Netherlands}
\altaffiltext{11}{Department of Astronomy, University of Wisconsin Madison, Madison WI, 53706, USA}
\altaffiltext{12}{Department of Astronomy, Yale University, New Haven, CT 06511, USA}

\begin{abstract}

We present the first results from a Hubble Space Telescope WFC3/IR program, which obtained direct imaging and grism observations of galaxies near quasar sightlines with a high frequency of uncorrelated foreground \ion{Mg}{2} absorption. These highly efficient observations targeted 54 \ion{Mg}{2} absorbers along the line of sight to nine quasars at $z_{qso}\sim2$. We find that $89$\% of the absorbers in the range $0.64< z < 1.6$ can be spectroscopically matched to at least one galaxy with an impact parameter less than 200 kpc and $|\Delta z|/(1+z)<0.006$. We have estimated the star formation rates and measured structural parameters for all detected galaxies with impact parameters in the range 7-200 kpc and star formation rates greater than 1.3 M$_{\odot}$ yr$^{-1}$.  We find that galaxies associated with \ion{Mg}{2} absorption have significantly higher mean star formation rates and marginally higher mean star formation rate surface densities compared to galaxies with no detected \ion{Mg}{2}.  Nearly half of the \ion{Mg}{2} absorbers match to more than one galaxy, and the mean equivalent width of the \ion{Mg}{2} absorption is found to be greater for groups, compared to isolated galaxies.  Additionally, we observe a significant redshift evolution in the physical extent of \ion{Mg}{2}-absorbing gas around galaxies and evidence of an enhancement of \ion{Mg}{2} within 50\degree~of the minor axis, characteristic of outflows, which persists to 80 kpc around the galaxies, in agreement with recent predictions from simulations.

\end{abstract}


\section{Introduction}

The in-situ evolution of galaxies is understood to be regulated by the accretion, consumption, heating, and expulsion of gas. Distant luminous quasars are uniquely powerful probes of these gaseous processes, as they enable the detection of intervening galaxies by virtue of their gas cross-sections, irrespective of their stellar luminosities, from the earliest epochs of galaxy formation to the present. 

The most prolific metal absorption transition in optical quasar spectra, singly-ionized magnesium (\ion{Mg}{2}), traces $T\sim10^{4}$ K photo-ionized gas in a wide range of environments in and around intervening galaxies.  Absorption from \ion{Mg}{2} is expected, in theory, to occur in sightlines probing the interstellar and circumgalactic medium of galaxies \citep[e.g.,][]{BS69, PW97}, large-scale star-formation driven outflows \citep[e.g.,][]{Nulsen98}, and gas being stripped or accreted within the extended halos of galaxies \citep[e.g.,][]{Kacprzak10, Stewart11}. 

Despite the potential utility of \ion{Mg}{2} for tracing this wide variety of physical processes that have been implicated in the evolution of galaxies, mapping the distribution and kinematics of \ion{Mg}{2} around galaxies has long posed a challenge. Due to the relative faintness of \ion{Mg}{2} host galaxies and their close angular proximity to a much brighter background quasar, the luminous counterparts at the redshifts where \ion{Mg}{2} is most easily detected at optical wavelengths ($z > 0.3$) have, until recently, been difficult to properly resolve from the ground.

\begin{table*}[!ht]
\centering
\begin{minipage}{0.95\textwidth}
\begin{center}
\caption{Quasar Targets\label{tbl-targets}}
\begin{tabular}{cccccccccc}
\tableline\tableline
Field & Name & RA & Dec & z$_{QSO}$ & SDSS  & SDSS  & SDSS & \ion{Mg}{2} \\
 &  & J2000 & J2000  &  &  Plate &  Fiber &  MJD & Absorbers \\
\tableline\tableline
\hline
0 & SDSS J001453.19+091217.6 &  3.7216917 & 9.20492  & 2.338 & 4536 & 770 & 55857 & 9 \\
1 & SDSS J082946.90+185222.0 & 127.44509 & 18.87293 & 1.792 & 2275 & 279  & 53709 & 5 \\
2 & SDSS J083852.05+025703.7 & 129.71690 & 2.95101  & 1.771 & 3809 & 414 & 55533 & 7 \\
3 & SDSS J091730.18+324105.5 & 139.37575 & 32.68483 & 2.017 & 1592 & 187 & 52990 & 4 \\
4 & SDSS J095432.63+354027.7 & 148.63599 & 35.67439 & 2.715 &  4573  & 140 &  55587 & 7\\
5 & SDSS J110742.74+102126.3 & 166.92809 & 10.35731 &  1.925 &   5361 &  900 & 55973 & 6 \\
6 & SDSS J113233.63+380346.4 & 173.14014 & 38.06290 & 2.302 & 4648 & 340 & 55673 & 5 \\
7 & SDSS J120342.24+102831.8 & 180.92601 & 10.47548 & 1.888  & 1228 & 556 & 52728 & 5 \\
8 & SDSS J120639.85+025308.3 & 181.66607 & 2.88564 & 2.518 & 4748 &  726 &  55631 & 6 \\
\tableline	
\end{tabular}
\end{center}
\end{minipage}
\end{table*}

In the past decade, the expansive spectroscopic quasar sample of the Sloan Digital Sky Survey (SDSS) has facilitated the detection of tens of thousands of \ion{Mg}{2} absorbers \citep[e.g.,][]{Nestor05, L09, Quider11, ZM13}.  Large samples have enabled statistical analyses, which circumvented many difficulties of directly detecting \ion{Mg}{2} hosts. Spatial clustering \citep[e.g.,][]{B06, Y06, L09, Gauthier09, Zhu2014, PR2015}, and stacking techniques \citep[e.g.,][]{Y06, Zibetti07, Menard08, NSM10, Menard11, Bordoloi11, Menard12} have provided statistical clues to the typical host galaxy properties as a function of \ion{Mg}{2} $\lambda\lambda$2796 rest-frame equivalent width (hereafter, W$_{r}$), which together indicate that the strongest absorbers (W$_{r} > 1$\AA) commonly trace dusty, large-scale outflows from star-forming galaxies.

However, these statistical correlations are not unequivocally supported by many observations at low redshift where the host galaxies can more easily be directly observed (e.g., Chen et al. 2010a); and despite their statistical links to vigorously star-forming galaxies, many galaxy hosts of high-W$_{r}$ \ion{Mg}{2} absorption remain undetected in even the deepest ground-based observations \citep{B07,B12a}, and some ultra-strong absorbers can be explained by the dynamics of gas in an intragroup medium \citep{Gauthier13}. Studies of directly detected host galaxies at $z < 0.5$ indicate that the origins of \ion{Mg}{2} are even less clear in the case of the more common population of absorbers with W$_{r} \leq 1$\AA, suggesting that \ion{Mg}{2} most frequently probes infalling gas from the halos of normal galaxies \citep{Chen10a,Chen10b, Kacprzak11a}. 

Recent advances in integral-field spectroscopy have enabled more efficient searches for the luminous counterparts of quasar absorption systems with considerably higher completeness. Surveys of quasar fields using VLT/MUSE \citep{Bacon10} have revealed that a single absorber can often be matched spectroscopically to multiple galaxies at the same redshift \citep[e.g.,][]{Bielby17, Peroux17, Klitsch18, Rahmani18, Peroux19, Hamanowicz20, Bielby2020, Dutta2020}, suggesting that multiple galaxies may contribute to intra-group gas \citep{Whiting06, Kacprzak10, Gauthier13} and further complicating the process of correlating galaxy and absorber properties.  These findings have called into question a long-standing simplifying assumption that an absorber can be assigned with high confidence to the nearest spectroscopically confirmed galaxy \citep[e.g.,][]{Bergeron88, BB91, Steidel94, LeBrun97}. Thus, the historic observational challenge of spectroscopically confirming even one candidate host galaxy for a particular absorber has now been replaced by the new challenge of determining which and how many galaxies with matching redshifts are truly associated with any single absorption system.

These recent advances in ground-based integral field spectroscopy, which have dramatically improved the efficiency in detecting the luminous counterparts of quasar absorbers, still lack the ability to resolve galaxies at small impact parameters ($<1.5\arcsec$) to the background quasar. This limitation is particularly problematic for studies of absorption-selected galaxies at high redshift. Space-based observations are thus still required in order to mitigate the possibility of absorber host galaxies being missed as a result of being blended with the point spread function (PSF) of the brighter quasar.

By analyzing observations of a single quasar from the AGHAST \citep{aghast} and 3D-HST surveys \citep{PvD11, Brammer12}, \citet{L12} provided a proof of concept that relatively shallow HST WFC3/IR G141 grism observations could be used to detect H$\alpha$ emission from the host galaxies of $1<z<2$ \ion{Mg}{2} absorbers. The three absorber-galaxy pairs identified by \citet{L12} had stellar masses consistent with estimates for the strong \ion{Mg}{2} absorber population from large-scale clustering analyses \citep{L11}.  The star formation rate surface densities of each galaxy were also estimated and found to exceed the threshold of local galaxies with large-scale outflows.  However, due to the small sample size, it was not possible to determine whether there was any particular azimuthal geometry to the \ion{Mg}{2}-absorbing circumgalactic medium, which could provide additional clues to the wind-driving capacity of the $z\sim1.5$ \ion{Mg}{2} host galaxies.

The survey described in this paper utilizes this same observational technique as \citet{L12}: harnessing the unique capabilities of the WFC3/IR camera and grism aboard the HST to measure the star formation rates (SFRs),  impact parameters, inclination angles and structural parameters for 54 \ion{Mg}{2}-selected galaxies near the epoch of peak star formation. Our targets are drawn from the largest multi-ion samples of quasar absorption lines compiled to date \citep{York21} from the SDSS Data Release 7 \citep{sdssdr7, Schneider10} and the SDSS-III Data Release 9 \citep{sdssdr9, Paris12}.  By mining these samples we have identified the nine SDSS quasar sightlines richest in $W_{r}>0.1$\AA~ \ion{Mg}{2} absorption. Despite the exceptionally high incidence of \ion{Mg}{2} in these spectra, the absorbers are all widely separated in redshift and physically uncorrelated; thus, their chance alignment is not expected to bias their physical properties or environments. Using WFC3/IR imaging and grism observations of these fields we have analyzed the properties of a complete magnitude-limited sample of \ion{Mg}{2}-selected galaxies at $z\geq 0.7$, where fewer than 100 \ion{Mg}{2} host galaxies have been spectroscopically confirmed to date. 

\begin{figure*}[!ht]
\centering
\begin{center}
\includegraphics[width=2.0\columnwidth]{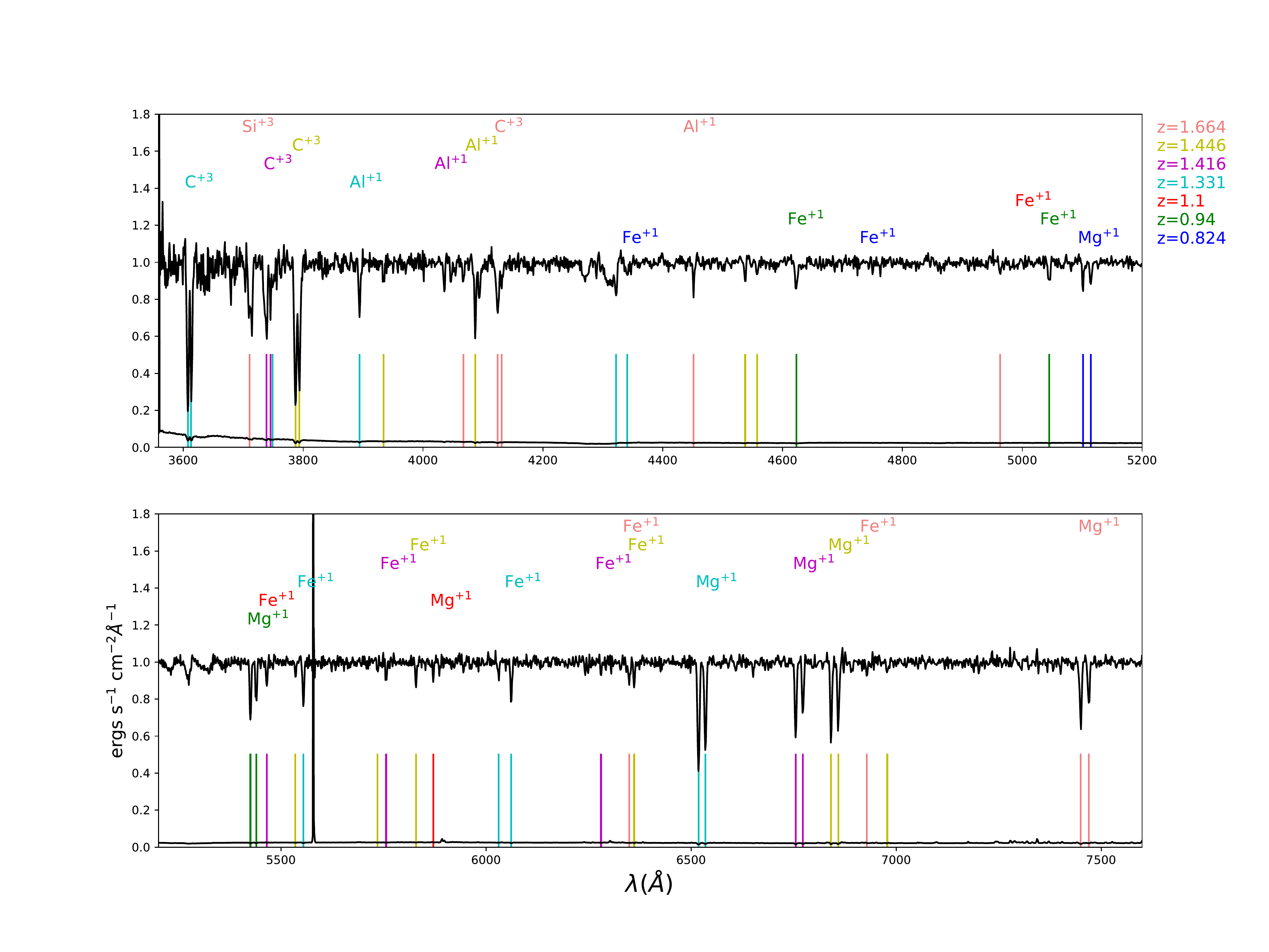}
\caption{\small An annotated portion of the  continuum-normalized SDSS BOSS spectrum of SDSS J083852.05+025703.7, one of the nine quasars targeted in this work. The uncertainty in the normalized flux per pixel is shown near the bottom of each panel. The signal to noise and the frequency and distribution of absorption in this spectrum is typical of the other eight targeted quasars, revealing seven strong \ion{Mg}{2} absorption systems in the range 0.6$<$z$<$1.7.  The most prominent individual ions detected in each absorption system are labeled in tiers with color-coding to match the redshifts provided to the right side of the top panel. Some identified absorption lines are not labelled in this figure due to space limitations. \label{fig:exspec}}
\end{center}
\end{figure*}

The observations contributing to this work are described in Section 2, and details of our analysis are given in Section 3.  In Section 4 we present a discussion of the results and their implications with regard to the evolving distribution of \ion{Mg}{2} around galaxies from $z\sim2$.   Throughout this paper, we assume a flat $\Lambda$--dominated CDM cosmology with $\Omega_m=0.3$, $H_0=70$ km s$^{-1} $Mpc$^{-1}$, and $\sigma_8=0.8$ unless otherwise stated. 
\vspace{0.5cm}

\section{Observations}

\subsection{Quasar Observations in the SDSS}

The quasar observations used in this analysis are drawn from the SDSS DR7 \citep{sdssdr7, Schneider10} and SDSS III DR9 \citep{sdssdr9, Paris12}. Each of the quasar spectra was run through an automated pipeline that detects strong ($W_{r}\gtrsim0.1\AA$) \ion{Mg}{2} absorption systems.  A detailed description of this process is provided in \citet{L09}. In brief, this pipeline extracts all $>3\sigma$ absorption features in continuum-normalized quasar spectra. The flux of each line is fit with a Gaussian profile to extract precise centroid and equivalent width measurements. In order to identify the ion and redshift corresponding to each measured absorption line, a line matching algorithm identifies pairs of 4$\sigma$ absorber detections with the correct wavelength separation expected for the doublet transitions of \ion{Mg}{2} ($\lambda\lambda$ 2796, 2803\AA) at a given redshift. The reliability of the identification is further quantified for each detected absorption system, taking into account the doublet ratio measured for \ion{Mg}{2}, the number of additional ions matched in absorption at the same redshift, and any blending with other absorption features identified at another redshift.  

From the catalogs extracted using this automated pipeline, we then identified all high signal-to-noise (SNR$>15$) quasar spectra with five or more physically uncorrelated and unambiguously intervening ($v/c>0.05$ in the quasar rest frame) \ion{Mg}{2} systems at $z>0.64$, where H$\alpha$ emission is accessible with the G141 grism.  All spectra meeting these criteria were then visually inspected to ensure the validity of the identified absorbers and the absence of additional complicating features such as broad absorption lines.  The sample of candidate quasars was then further reduced by removing objects close to bright nearby stars, which would saturate the {\it HST} observations.  We also required that the quasars be fainter than $m_{i}=17$ to facilitate the detection of galaxies in close angular proximity to the background quasar. 

Our final sample is composed of the nine quasars listed in Table ~\ref{tbl-targets}.  The selected quasar spectra exhibit a total of 44 high-confidence \ion{Mg}{2} absorbers in the redshift range accessible by H$\alpha$ in G141 ($0.64 < z \leq 1.6$), spanning an equivalent width range of $0.2 < W_{r} < 2.5$\AA. These spectra additionally include 12 absorption systems at $1.6 < z < 2.5$, whose host galaxies could also be determined by the detection of O[III]/H$\beta$ emission in the G141 grism observations. 

The catalog of absorption systems targeted by this program is presented in Table ~\ref{tbl-absorbers}, which includes redshifts and observed-frame equivalent width measurements for all identified \ion{Mg}{2} doublets.  Absorption systems that report a measurement for only the $\lambda2796$\AA~ \ion{Mg}{2} transition indicate cases where corresponding absorption at the wavelength of the weaker $\lambda2803$\AA~ transition fell below our 3$\sigma$ detection threshold.  Some of the higher redshift absorption systems listed in Table ~\ref{tbl-targets} lack equivalent width measurements for both transitions of \ion{Mg}{2}; these are cases in which an absorption system was detected by the presence of a \ion{C}{4} ($\lambda\lambda1548,1550$\AA) doublet, and either the \ion{Mg}{2} doublet was inaccessible due to the wavelength limitations of the SDSS spectrum, or both lines were undetected at the 3$\sigma$ level. Such systems were not included in our galaxy-absorber pair analysis but are documented for completeness.  Any detected absorption systems that have low confidence due to line blending, multiple degenerate line identifications, or having too few matching lines to provide an unambiguous redshift identification are flagged in the table and have been excluded from further analysis. Additionally, galaxies with poorly determined morphologies did not produce reliable azimuthal angles, and we did not attempt to find the morphologies of galaxies with uncertain grism redshifts.


\begin{figure}[t!]
\includegraphics[width=1.0\columnwidth]{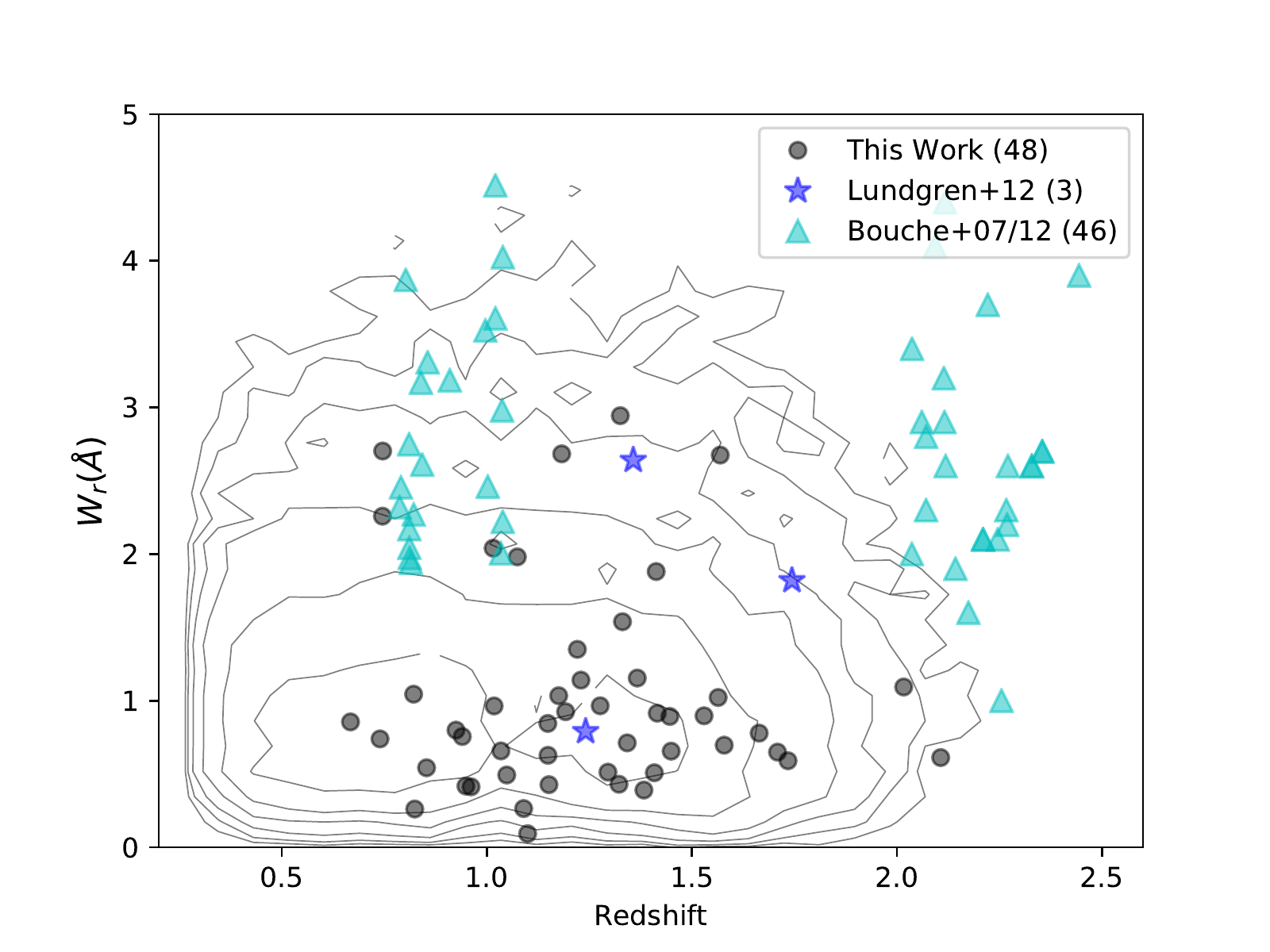}
\caption{\small The redshifts and rest-frame equivalent widths (W$_{r}$) of the $\lambda 2796$\AA~ transition for \ion{Mg}{2} absorbers targeted in this work, compared to \ion{Mg}{2} host galaxy studies at similar depths from \citet{B07}, \citet{B12a}, and \citet{L12}.  The parent distribution of the $\sim30,000$ absorbers in the SDSS sample is shown with contours, demonstrating that  our sample targets both a larger dynamic range in W$_{r}$ than previous high-redshift studies and more accurately represents the global distribution of \ion{Mg}{2} observed in medium resolution spectroscopic quasar surveys.  \label{fig:absdist}}
\end{figure}

\begin{figure}[t!]
\includegraphics[width=1.0\columnwidth]{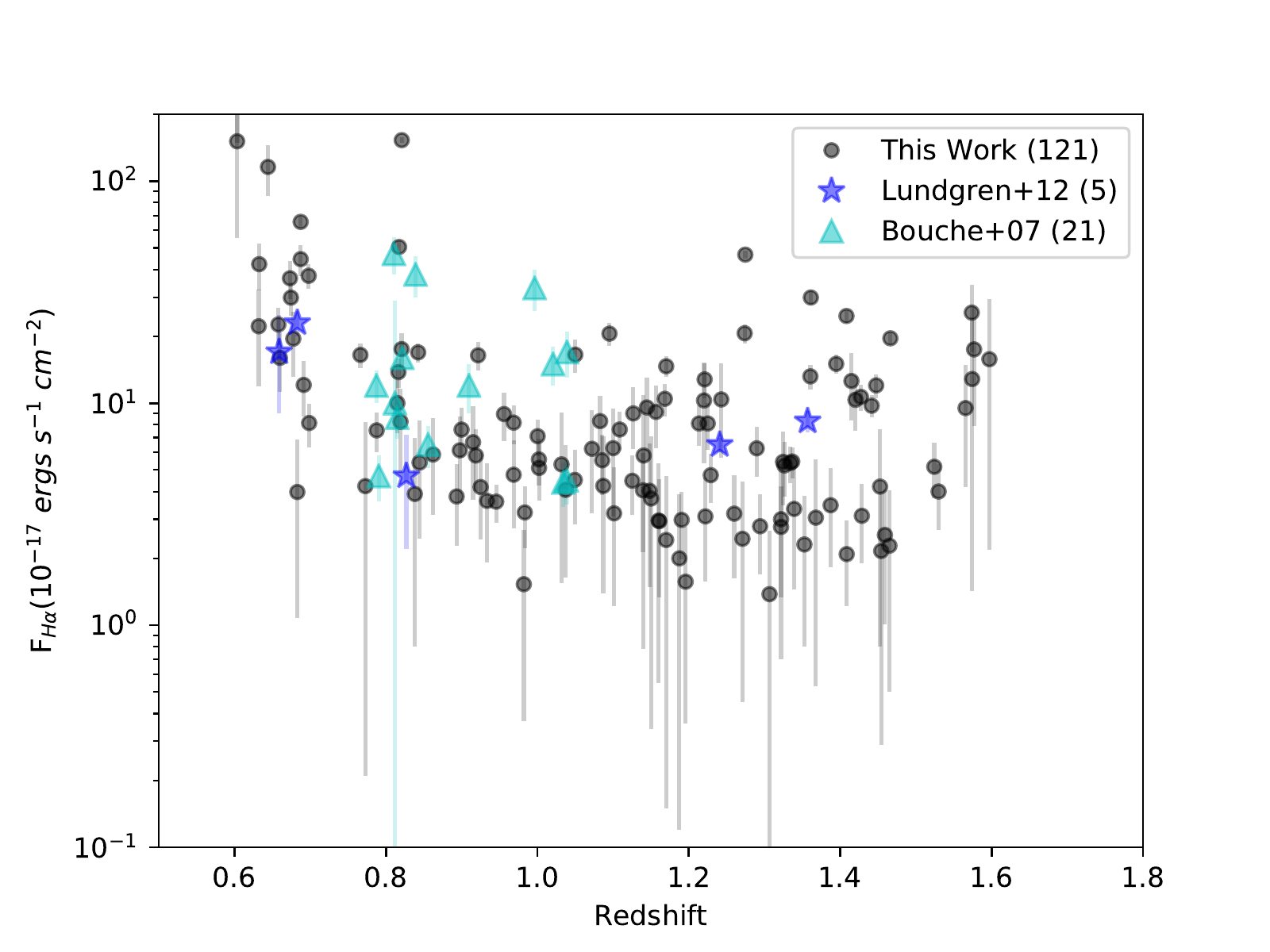}
\caption{\small H$\alpha$ flux versus redshift for all the galaxies with high-quality redshifts detected within 20\arcsec~of each quasar targeted in this work. For comparison, measurements from other \ion{Mg}{2} host galaxy surveys in this redshift range (\citet{B07}, \citet{B12a}, and \citet{L12}) are overplotted.  This figure includes all galaxies with reported H$\alpha$ flux measurements in the included samples, regardless of whether they were matched to a detection of \ion{Mg}{2} absorption. \label{fig:Hadist}}
\end{figure}

A typical quasar spectrum from our sample is shown in Figure ~\ref{fig:exspec}. The SDSS quasar spectra for sightlines targeted in this survey have a spectral resolution (R$\sim$1800) and mean signal-to-noise ratio comparable to that of the Keck DEIMOS spectrum from our pilot study \citep{L12}. In Figure ~\ref{fig:absdist} we compare the rest-frame equivalent width (W$_{r}$) and redshift distribution of the \ion{Mg}{2} absorbers in our sample to that of other surveys of \ion{Mg}{2} host galaxies at similar sensitivities. The global distribution of the $\sim30,000$ \ion{Mg}{2} absorbers detected in the SDSS DR7 quasar catalog of \citet{York21} is shown in contours, demonstrating that our targets sample a larger dynamic range in W$_{r}$ and also more accurately represent the global population of \ion{Mg}{2} absorbers detected in medium resolution spectroscopic quasar surveys.  Figure ~\ref{fig:Hadist} illustrates the H$\alpha$ flux limit for galaxies detected in this work, compared to other studies of \ion{Mg}{2}-absorbing galaxies at similar redshifts.

\subsection{HST WFC3/IR Observations}

Following a similar strategy to the AGHAST Survey \citep{aghast}, the 600 arcmin$^{2}$ 3D-HST Survey \citep{PvD11, Brammer12} employed {\it HST}/WFC3 G141 grism observations, paired with {\it HST}/WFC3 F140W direct imaging, to extract 2-dimensional spectra for $\sim$80,000 objects to a 5$\sigma$ limiting depth of m$_{H160}\sim26$ and produce a galaxy redshift catalog comparable to the deepest ground-based surveys \citep{Momcheva16}.  As shown in \citet{L12}, which leveraged the observations and data products from AGHAST and 3D-HST, WFC3/IR grism observations also enable the efficient detection and redshift confirmation of galaxies at $1<z<2$ with small impact parameters ($\gtrsim 7$ kpc) to quasars. Our program has adopted this well-tested approach in order to build up a relatively large sample of high-redshift galaxy-absorber pairs, with the aim of better understanding the proper-
\FloatBarrier

\begin{table*}[h!]
\centering
\begin{minipage}{0.95\textwidth}
\begin{center}
\caption{Identified Absorption Systems and Proximate Galaxy Candidates\label{tbl-absorbers}}

\begin{tabular}{cccccccccccc}

\tableline\tableline
Target &  $z_{abs}$ & $W_{obs}^{2796}$ & $W_{obs}^{2803}$ &  flag\footnote{Absorber quality flag; 1 = system identification uncertain} & $z_{gal}$ & $\rho$ & $\phi$ & F$_{H\alpha}$ & flag\footnote{Galaxy redshift quality flag; 1 = uncertain} & flag\footnote{Matching flag; 1 = galaxy redshift matches the \ion{Mg}{2} redshift, within the allowed limits.}\\
& & [\AA] & [\AA] & & & [kpc] & [deg] &  [10$^{-17}$~erg~s$^{-1}$~cm$^{-2}$] & & &\\
\tableline\tableline
\hline
0 & 0.66759 & $1.426\pm0.072$ & $1.952 \pm 0.112$ & 0 & 0.673 & 67.7 & 18.7 & $36.5\pm7.2$  & 0 & 1\\
& & & & & 0.678 & 81.9 & 58.4 & $19.5\pm6.4$ & 0 & 0 \\ 
 
&1.14918&$1.348\pm0.106$&$1.208\pm0.116$ & 0 & 1.145 & 74.2 & 33.9 & $9.6\pm3.5$  & 0 & 1\\
&  & & & & 1.148 & 144.4 & -- &$1.14\pm5.72$ & 1 & 0\\

&1.17525&$2.249\pm0.11$&$1.796\pm0.107$&0 & 1.160 & 121.1 & -- &$0.6\pm1.2$ & 0 & 0\\ 
& & & & & 1.161 & 135.4 & 68.9 &$2.9\pm1.6$ & 0 & 0\\ 
& & & & & 1.170 & 190.2 & 87.9 &$2.50\pm2.3$ & 0 & 0\\
 
&1.22071&$2.998\pm0.11$&$2.172\pm0.102$&0 & 1.236 & 56.5 & -- &$1.0\pm3.2$ & 1 & 0\\
& & & & & 1.239 & 57.6 & -- &$0.0\pm0.1$ & 1 & 0\\
& & & & & 1.243 & 91.2 & 77.7 &$10.4\pm4.7$ & 0 & 0\\ 
& & & & & 1.222 & 96.2 & 37.3 &$3.1\pm1.5$ & 0 & 1\\
& & & & & 1.221 & 174.5 & 31.7 &$12.8\pm2.3$ & 0 & 0\\
& & & & & 1.214 & 178.8 & 8.82 &$8.1\pm1.7$ & 0 & 0\\

&1.52965&$2.269\pm0.117$&$1.449\pm0.112$&0 & 1.524 & 99.8 & 70.4 &$4.7\pm1.4$ & 0 & 1\\

& 1.56403 & $2.620\pm0.099$ & -- & 0 & 1.574 & 38.0 & 43.8 &$12.8\pm11.4$ & 0 & 0\\
& & & & & 1.577 & 49.8 & 79.9 &$17.5\pm9.5$ & 0 & 0\\
& & & & & 1.574 & 93.5 & 28.8 &$25.6\pm8.6$ & 0 & 0\\

&1.56916&$6.873\pm0.137$&$3.631\pm0.126$&0 & 1.574 & 38.0 & 43.8 & $12.8\pm11.4$ & 0 & 1\\
& & & & & 1.577 & 49.8 & 79.9 &$5.1\pm2.9$ & 0 & 1\\
& & & & & 1.574 & 93.5 & 28.8 &$25.6\pm8.6$ & 0 & 1\\

&2.13426&$0.898\pm0.173$&-- &0 & --&-- &-- &--\\
&2.22914&$0.737\pm0.083$&-- &0 & --&-- &-- &--\\

\hline
1&0.85303&$1.005\pm0.091$&$1.151\pm0.129$&0 & 0.844 & 63.9 & 49.1 & $5.4\pm3.0$ & 0 & 1\\
& & & & & 0.857 & 88.7 & -- & $0.0\pm0.0$ & 1 & 0\\
& & & & & 0.843 & 150.3 & -- & $1.0\pm8.8$ & 1 & 0\\

&1.01773&$1.945\pm0.183$&$0.969\pm0.14$&0 & --&-- &-- &-- &--\\

& 1.36680 & $2.732\pm0.144$ & $2.574\pm0.149$ &0 & 1.361 & 65.0 & 56.5 & $29.9\pm2.2$ & 0 & 1 \\
& & & & & 1.353 & 65.6 & -- & $2.3\pm1.5$ & 0 & 1\\
& & & & & 1.368 & 132.2 & 79.3 & $3.0\pm2.5$ & 0 & 1 \\

&1.40842&$1.224\pm0.102$&$1.073\pm0.101$&0  & 1.407 & 84.5 & 54.1 & $0.0\pm0.1$ & 0 & 1\\
&1.80016&$4.243\pm0.177$&$3.628\pm0.184$&0 & --&-- &-- &-- & --\\

\hline

2&0.82426&$0.476\pm0.058$&$0.162\pm0.036$ & 0 & 0.821 & 137.8 & 42.7 & $16.6\pm3.0$ & 0 & 1\\
& & & & & 0.817 & 166.1 & -- & $13.8\pm2.1$ & 0 & 0\\ 

&0.94003&$1.474\pm0.070$&   --  &0 & 0.934 & 102.7 & -- & $3.6\pm1.7$ & 0 & 1\\ 

&1.09951&$0.194\pm0.035$&   --  &0  & 1.095 & 60.6 & -- & $22.0\pm2.7$ & 0 & 1\\
& & & & & 1.109 & 191.6 & 79.6 & $7.6\pm1.5$ & 0 & 0\\

&1.33095&$3.624\pm0.067$&$2.787\pm0.074$&0 & 1.325 & 40.0 & 51.9 & $5.4\pm2.0$ & 0 & 1\\
& & & & & 1.318 & 112.1 & 49.7 & $1.3\pm16.1$ & 0 & 1 \\
& & & & & 1.347 & 173.0 & 43.7 & $1.8\pm2.1$ & 0 & 0\\

&1.41576&$2.240\pm0.067$&$1.448\pm0.071$&0 & 1.395 & 93.2 & -- & $15.0\pm1.5$ & 0 & 0\\ 
& & & & & 1.409 & 135.3 & 59.6 & $2.1\pm0.9$ & 0 & 1\\

&1.44630&$2.182\pm0.072$&$1.965\pm0.07$&0 & 1.448 & 87.8 & -- & $12.0\pm1.5$ & 0 & 1\\
& & & & & 1.453 & 127.4 & -- & $4.2\pm3.4$ & 0 & 1 \\ 

&1.66407&$2.074\pm0.086$&$1.334\pm0.089$&0 & -- & -- & -- & -- & --\\
\hline

3&1.22947&$2.542\pm0.202$&$2.536\pm0.164$&0& 1.225 & 39.6 & 69.2 & $8.1\pm1.9$ & 0 & 1\\
& & & & & 1.220 & 43.7 & 67.0 & $10.3\pm4.9$ & 0 & 1\\
& & & & & 1.229 & 87.5 & 67.3 & $4.7\pm1.2$  & 0 & 1\\
& & & & & 1.214 & 180.3 & -- & $0.0\pm0.0$ & 1 & 0\\

&1.41294&$4.538\pm0.176$&$4.253\pm0.186$&0 & 1.415 & 15.7 & 51.8 & $13.9\pm4.5$ & 0 & 1\\ 
& & & & & 1.408 & 66.0 & 77.3 & $24.7\pm1.6$ & 0 & 1\\
& & & & & 1.413 & 166.4 & 26.9 & $0.7\pm1.3$ & 0 & 0\\

&1.45627&$2.043\pm0.338$  &  --   &1 & --&-- &-- &-- &--\\
& 1.53688 & -- & -- & 0 & --&-- &-- &-- &--\\
&2.01657&$3.298\pm0.674$&$1.897\pm0.316$&0& --&-- &-- &-- &--\\
\hline

4&1.07438&$4.109\pm0.188$&$3.36\pm0.204$&0  & 1.072 & 19.3 & -- & $6.2\pm3.0$ & 0 & 1\\

&1.18256&$5.857\pm0.276$&$5.128\pm0.268$&0 & 1.188 & 120.1 & 4.1 & $2.0\pm1.9$ & 0 & 1\\
& & & & & 1.191 & 156.6 & 45.5 & $3.0\pm1.0$ & 0 & 0\\
&1.35044&$0.727\pm0.111$&$0.756\pm0.201$&0& --&-- &-- &-- &--\\
&1.44949&$1.606\pm0.182$&$1.219\pm0.211$&0& 1.442 & 63.9 & 44.5 & $9.7\pm1.1$ & 0 & 1\\
& & & & & 1.427 & 154.6 & 37.2 & $10.7\pm1.5$ & 0 & 0\\ 

&1.70888&$1.758\pm0.222$&$1.728\pm0.217$&0 &--&-- &-- &-- &--\\
&1.73440&$1.614\pm0.199$&$1.457\pm0.187$&0 &--&-- &-- &-- &--\\
&2.01760& -- & -- & 0& --&-- &-- &-- &--\\
&2.10673&$1.901\pm0.281$&$0.778\pm0.175$&0& --&-- &-- &-- &--\\

\tableline	
\end{tabular}
\end{center}
\end{minipage}
\end{table*}
\begin{table*}
\centering
\begin{minipage}{0.95\textwidth}
\begin{center}
\caption{Identified Absorption Systems and Proximate Galaxy Candidates (continued)\label{tbl-absorbers2}}
\begin{tabular}{cccccccccccc}
\tableline\tableline
Target &  $z_{abs}$ & $W_{obs}^{2796}$ & $W_{obs}^{2803}$ &  flag\footnote{Absorber quality flag; 1 = system identification uncertain} & $z_{gal}$ & $\rho$ & $\phi$ & F$_{H\alpha}$ &  flag\footnote{Galaxy redshift quality flag; 1 = uncertain} & flag\footnote{Matching information; 1 = galaxy redshift matches the \ion{Mg}{2} redshift, within the allowed limits.}\\
& & [\AA] & [\AA] & & & [kpc] & [deg] & [10$^{-17}$~erg~s$^{-1}$~cm$^{-2}$] & & & \\
\tableline\tableline
\hline
& & & & &  & &  & \\
5&0.74548&$3.942\pm0.056$&$2.888\pm0.055$&0& 0.741 & 35.4 & -- & $0.0\pm0.0$  & 0 & 1\\
& & & & & 0.735 & 40.0 & -- & $0.1\pm1.4$ &  1 & 0\\
& & & & &0.742 & 113.3 & -- & $71.2\pm17.1$ &  0 & 1\\

&1.01540&$4.111\pm0.078$&$1.978\pm0.049$&0 & 1.011 & 43.5 & -- & $10.0\pm1.9$  & 0 & 1\\
& & & & & 1.008 & 81.2 & -- & $5.9\pm3.2$ & 0 & 1\\

&1.04852&$1.010\pm0.065$&$0.907\pm0.072$&0& -- &-- &--&-- &--\\

&1.14783&$1.815\pm0.069$&$1.114\pm0.067$&0& --&-- &-- &-- &--\\ 
& 1.29529 & $1.176\pm0.081$ & $0.742\pm0.081$ & 0 & 1.295 & 21.7 & -- & $0.0\pm0.0$ & 1 & 0\\
& & & & &1.277 & 56.8 & -- & $8.4\pm3.7$ & 1 & 0\\
& & & & & 1.289 & 58.3 & -- & $18.3\pm2.6$ & 0 & 1\\
& & & & &1.291 & 74.0 & -- & $15.2\pm4.7$ & 0 & 1\\

&1.32544&$6.846\pm0.067$&$6.114\pm0.053$&0& 1.325 & 14.7 & -- & $0.0\pm0.0$ & 1 & 0\\
 & & & & & 1.336 & 69.7 & -- & $10.1\pm1.6$ & 0 & 1\\
 & & & & &1.340 & 159.0 & -- & $8.6\pm1.6$ & 0 & 0 \\
& 1.60117 & -- & -- & 0 & --&-- &-- &-- &--\\
& 1.71102 & -- & -- & 0 & --&-- &-- &-- &--\\
& 1.86376  & -- & -- & 0 & --&-- &-- &-- &--\\
& 1.88273 & -- & -- & 0 & --&-- &-- &-- &--\\
& & & & &  & &  & \\
\hline
& & & & &  & &  & \\
6&0.92480&$1.538\pm0.084$&$1.265\pm0.086$&0& 0.919 & 40.8 & 54.9 & $5.8\pm1.8$ & 0 & 1\\
 & & & & & 0.922 & 142.9 & 0.0 & $16.4\pm2.4$ &0 & 1\\
 & & & & & 0.915 & 176.8 & 82.9 & $6.7\pm3.0$ & 0 & 0\\
&1.15147&$0.919\pm0.119$&$0.604\pm0.108$&0& --&-- &-- &-- &--\\

&1.19213&$2.027\pm0.106$&$1.873\pm0.107$&0& --&-- &-- &-- &--\\
&1.27638&$2.197\pm0.113$&$2.152\pm0.112$&0& 1.275 & 86.2 & 70.2 & $46.6\pm2.0$ & 0 & 1\\
 & & & & & 1.281 & 158.5 & -- & $1.5\pm8.7$ & 1 & 0\\
& & & & & 1.274 & 171.0 & 49.1 & $20.6\pm2.1$ & 0 & 0\\
&1.38301&$0.930\pm0.109$&$0.633\pm0.124$&0& 1.389 & 55.7 & 71.2 & $1.2\pm5.5$ & 0 & 1\\
& 1.75775 & -- & -- & 1 & --&-- &-- &-- &--\\
& 2.07231 & -- & -- & 0 & --&-- &-- &-- &--\\
& 2.28424  & -- & -- & 0 & --&-- &-- &-- &--\\
& & & & &  & &  & \\
\hline
& & & & &  & &  & \\
7&0.73955&$1.286\pm0.115$&$0.878\pm0.116$&0& 0.740 & 39.1 & 12.4 & $0.0\pm 0.0$ & 0 & 1\\
& & & & &0.745 & 71.5 & 69.0 & $1.5\pm14.6$ & 1 & 0 \\
&0.74595&$4.719\pm0.137$&$3.935\pm0.132$&0 & 0.740 & 39.1 & 12.4 & $0.0\pm0.0$  & 0 & 0\\
 & & & & & 0.745 & 71.5 & 69.0 & $1.5\pm14.6$  & 0 & 1\\

&1.32224&$0.998\pm0.129$&$0.611\pm0.102$&0& 1.334 & 92.3 & 55.4 & $5.4\pm1.0$ & 0 & 1\\
& & & & & 1.339 & 95.7 & 35.7 & $1.7\pm3.1$  & 0 & 0\\ 
&1.34230&$1.668\pm0.125$&$1.234\pm0.128$&0& 1.334 & 92.3 & 55.4 & $5.0\pm2.1$  & 0 & 0\\
 & & & & & 1.339 & 95.7 & 35.7 & $1.7\pm3.1$  & 0 & 1\\
& 1.57449  & -- & -- & 1 & --&-- &-- &-- &-- &--\\  
&1.57876&$1.796\pm0.126$&$1.472\pm0.137$&0 & 1.566 & 40.4 & 83.8 & $9.5\pm5.3$  & 0 & 1\\
& & & & &1.589 & 56.2 & -- & $19.1\pm28.4$  & 1 & 0\\
& 1.79606  & -- & -- & 0 & --&-- &-- &-- &-- &--\\ 
& & & & &  & &  & \\
\hline
& & & & &  & &  & \\
8&0.68079&$1.309\pm0.084$&$0.477\pm0.069$&1 & 0.675 & 33.4 & -- & $29.9\pm5.0$  & 0 & 0\\ 
 & & & & & 0.691 & 118.4 & -- & $12.1\pm3.4$  & 0 & 0\\
 & & & & & 0.686 & 131.2 & -- & $1.9\pm4.2$  & 1 & 0\\
&0.82150&$1.902\pm0.109$&$1.386\pm0.102$&0 & 0.815 & 65.0 & 66.5 & $9.9\pm2.7$ & 0 & 1\\
& & & & &0.819 & 85.7 & 25.6 & $8.2\pm3.4$  & 0 & 1\\
 & & & & &0.812 & 99.7 & 71.6 & $0.2\pm0.8$  & 0 & 1\\
 & & & & &0.817 & 116.4 & 18.1 & $50.5\pm3.8$  & 0 & 1\\
&0.94898&$0.814\pm0.075$&$0.408\pm0.067$&0 & 0.956 & 36.9 & 65.2 & $8.9\pm2.2$  & 0 & 1\\
 & & & & &0.946 & 76.6 & 9.4 & $3.6\pm0.7$ &  0 & 1\\
&0.96149&$0.809\pm0.078$&$0.45\pm0.086$&0 & 0.956 & 37.0 & 65.2 & $8.9\pm2.2$  & 0 & 0\\
&1.08980&$0.552\pm0.101$&$0.643\pm0.116$&0& 1.086 & 82.0 & 53.2 & $5.5\pm1.7$  & 0 & 1\\
&1.36595&$1.110\pm0.127$&$0.781\pm0.147$&0& 1.361 & 60.4 & 56.5 & $13.2\pm1.7$  & 0 & 1\\
 & & & & &1.376 & 127.6 & -- & $1.6\pm2.3$ & 1 & 0\\
 & & & & &  & &  & \\
\tableline	
\end{tabular}
\end{center}

\end{minipage}
\end{table*}
\FloatBarrier

\begin{figure*}[!t]
\centering
\begin{center}
\includegraphics[width=1.95\columnwidth]{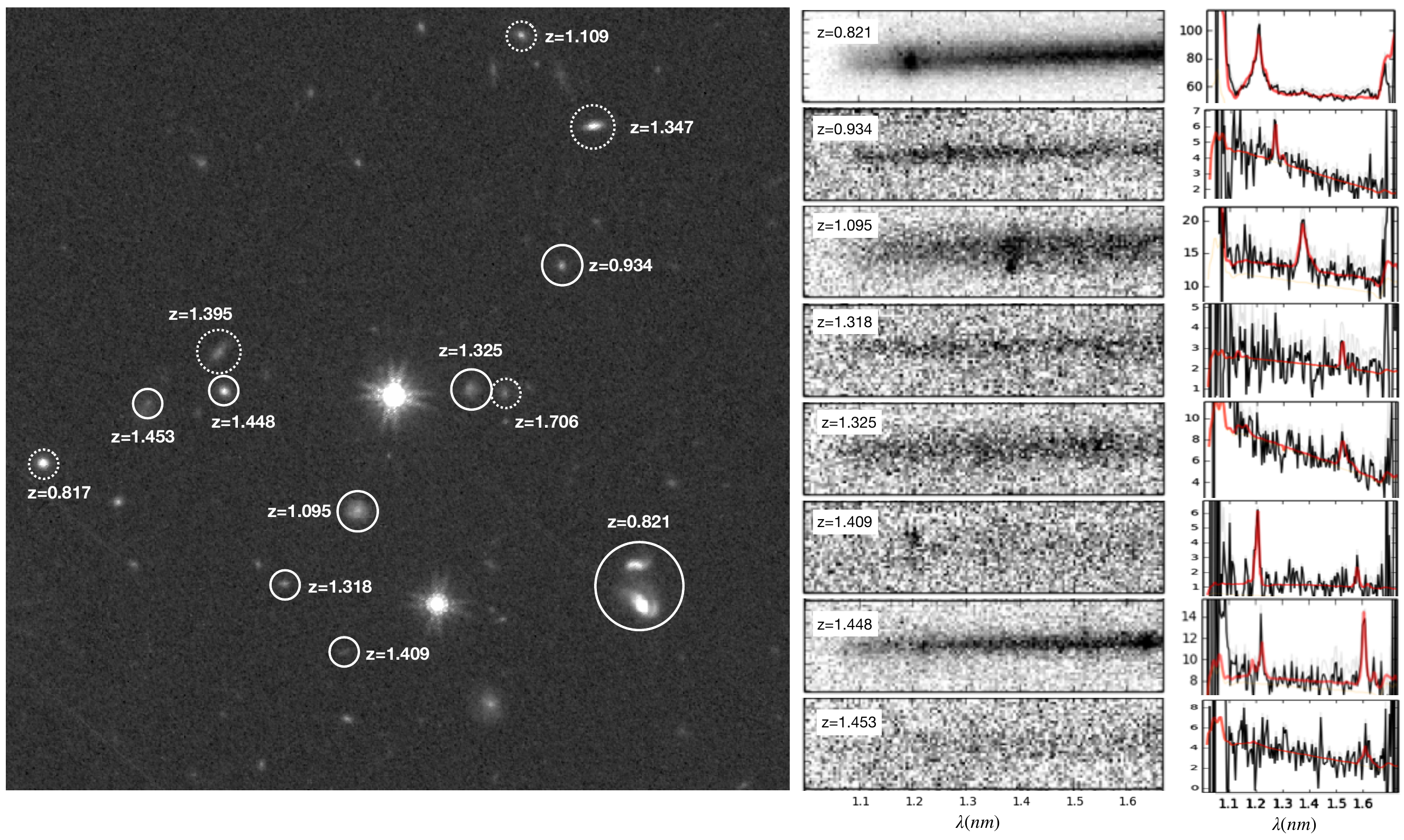}
\caption{\small WFC3/IR observations of galaxies with redshifts matching six strong \ion{Mg}{2} absorption systems detected in the spectrum of quasar SDSS J083852.05+025703.7 (field 2, $z_{QSO}=1.771$). Left: The F140W direct image of the field centered on the quasar, cropped to a field of view of 0.8'$\times$0.8'. The nearest galaxies with well-determined redshifts are labeled. Solid circles indicate galaxies matched to absorption; dashed circles indicate galaxies with well-determined redshifts but are not considered a match to absorption due to the redshift difference. The 2D (middle panel) and 1D (right panel) G141 grism spectra for the  absorption-matched galaxies labeled in the image of the field. In the 1D spectra, observations are presented in black with the best-fit model overlaid in red. The H$\alpha$ emission line is clearly visible for the galaxies with redshifts below $z=1.6$. Only one of the two acquired grism spectra for the pair of interacting galaxies at z=0.821 is shown, but both spectra indicate the same redshift with high signal to noise. \label{fig:absmatch}}
\end{center}
\end{figure*}

\noindent ties of absorption-selected galaxies at these redshifts.  

In an 18-orbit Cycle 21 {\it HST} general observing program (GO-13482, PI Lundgren) we obtained four dithered G141 grism exposures for each quasar target along with a short (4 x 200s) direct image in F140W, which is necessary for target identification, wavelength and flux calibration, and for characterizing the structural parameters of the absorber host galaxies. Each two-orbit visit consisted of a guide star acquisition and re-acquisition and four pairs of dithered (with half pixel offsets) 200s F140W images and 1150s G141 grism exposures.

Given our selection of quasars with a high incidence of intervening absorption, we anticipated that contamination from many galaxies in close proximity to the quasar on the sky could be a complicating factor. Thus, we applied a random position angle offset between each of the two G141 visits, in order to minimize the potential effects of contamination from the overlapping spectra of closely spaced galaxies.  These orientation offsets enabled us to combine the full G141 data for maximum signal to noise for clearly observed galaxy grism spectra, and to extract spectral information from individual orientations for galaxies that were catastrophically contaminated by another bright nearby object in an individual orientation.

We obtained F140W direct imaging and G141 spectra for galaxies in the targeted fields to a limiting magnitude of m$_{F140W} = 24$ and a limiting 3$\sigma$ H$\alpha$ flux limit of $3\times10^{17}$ ergs s$^{-1}$ cm$^{-2}$ (equivalent to a SFR $> 1.3$ M$_{\odot}$ yr$^{-1}$ at $z = 1$, assuming a Salpeter initial mass function) and at impact parameters around the central quasar limited only by the quasar PSF ($\gtrsim 5$ kpc) and the field of view ($\lesssim$ 480 kpc at $z=1$).  The G141 grism observations provided slitless spectroscopy over the wavelength range 1.10$-$1.65$\mu$m with a first-order dispersion of 46.5 \AA/pixel (R $\sim130$) and a spatial resolution of $\sim0.\arcsec13$, sampled with 0.\arcsec06 pixels.   These specifications enabled the detection of H$\alpha$ emission in the redshift range $0.64<z<1.6$, [\ion{O}{3}]$\lambda$5007 in the range $1.2<z<2.3$, and [\ion{O}{2}]$\lambda$3727 for $2.0<z<3.4$.

\section{Analysis}

\subsection{Analysis of Foreground Galaxies}

The WFC3/IR galaxy observations were reduced and analyzed using a variation on the method described in \citet{Brammer12}.  This process relies on first detecting sources in the F140W images using \textit{SExtractor} \citep{BA96} and then extracting the individual G141 spectra for each source, minus the modeled contamination from the sky background and any nearby objects.  A careful visual inspection was also conducted for all of the extracted 2D grism spectra for galaxies within 24\arcsec ($\sim200$ kpc at z=1.2) of the central quasar in each field. During this process, badly contaminated grism orientations were vetoed from contributing to the spectral reduction.  For most galaxies, the contamination of the grism data was minimal in both orientations, and the two observations could be combined to achieve the highest possible signal-to-noise.

We identified all galaxies above our detection limit with $0.64<z<2.6$ in the surrounding field.  The applied redshift limitations correspond to the range in which emission from either H$\alpha$ or the [\ion{O}{3}]/H$\beta$ complex is observable in the G141.  Although the 5$\sigma$ depth of the F140W images extends to $m_{140}\sim26$, the continuum limit at which grism redshifts may be reliably obtained for galaxies at $1<z<3$ in the 3D-HST Survey effectively limits our search to galaxies with $m_{140}\lesssim24$ \citep{Brammer12}.

In contrast to the pilot study of \citet{L12}, the target fields in this work do not benefit from supplementary deep archival multi-wavelength imaging from which to generate SED fits and photometric redshift priors. Thus, the redshift determinations for galaxies in these fields are entirely dependent on the quality and features of the G141 grism data.  The highest confidence galaxy redshifts are determined from G141 spectra with multiple strong emission lines.  Galaxies without emission lines (e.g., quiescent galaxies) generally did not allow for conclusive redshift determinations unless the continuum signal-to-noise ratio (SNR) was high. 

Based on the findings of \citet{Brammer12} and \citet{Momcheva16}, we estimate the galaxy redshifts in this work have a typical precision of $|\Delta z| / (1+z)=0.0035$ at $z>0.7$.  However, the precision is expected to increase with the strength and number of emission lines in each G141 spectrum. \citet{Bielby2019}, which compared spectroscopic redshifts of galaxies with 3-hour VLT/MUSE observations and eight orbits of coverage with HST/WFC3 G141, reported that the R=130 spectral resolution of the G141 grism results in a typical velocity accuracy of $\sigma_{v}=682$ km s$^{-1}$ at $z=1$.


For each of the galaxies with H$\alpha$ emission line measurements in the G141 spectra we estimate the SFR using the relation from \citet{Kennicutt98}:
\begin{equation}\label{eq:SFR}
\text{SFR}_{H\alpha}~[\text{M}_{\odot}~\text{yr}^{-1}] = \frac{L_{H\alpha}}{ 7.9 \times 10^{-42}~[\text{erg s}^{-1}]}
\label{eqn-1}
\end{equation}
which assumes the initial mass function of \citet{Salpeter55}. Given to the wavelength range of the grism data, the H$\alpha$ and H$\beta$ emission lines are not concurrently accessible for the galaxies with $z<1.2$. Subsequently, we have not been able to apply a standard correction for dust extinction to the H$\alpha$ flux measurements in our full sample, and these estimated SFRs represent lower limits. We also note that at the resolution of the WFC3/IR G141 grism spectroscopy, H$\alpha$ and N[II] emission are unresolved. Thus, we expect that the H$\alpha$ fluxes are also over-estimated on the order of 25-33\% \citep{Villar2008,Geach2008,Sobral2009}.

For galaxies in which emission lines are not evident or observable in the G141 1D spectra, ancillary multi-band photometric observations are required to precisely estimate the stellar masses and SFRs with stellar population synthesis (SPS) modeling.  Future follow-up multi-wavelength HST imaging of these fields would add significant value to this survey and will be pursued in upcoming cycles.

\subsubsection{Matching Galaxies to \ion{Mg}{2} Absorption}

In searching for galaxies matching to the \ion{Mg}{2} absorbers in our targeted quasars, we have limited our analysis to the redshift range in which H$\alpha$ is observable in the G141 grism data ($0.64\leq z\leq1.6$).  These limits provide a homogeneous sample for which the star formation rates can be determined in a self-similar way. We also include in our search only galaxies with apparent magnitudes of $m_{140}<25$ and high-confidence redshifts, as determined through careful visual inspection of the extracted grism spectrum for each galaxy.  We required that the galaxies match in redshift to \ion{Mg}{2} within $|\Delta z|/(1+z_{MgII})<0.006$, equivalent to $\Delta v<900$ km s$^{-1}$ at $z=1$, and have an impact parameter less than 200 kpc. The redshifts of the quasar absorption line systems have been precisely determined to five decimal places by taking the average redshift of the individually modeled centroids for each 3$\sigma$ absorption line identified in each system. Thus, the uncertainty in matching galaxies to the \ion{Mg}{2} systems lies exclusively in the galaxy redshift estimation from the grism data.  Resolved galaxies matching the aforementioned criteria whose redshifts were determined to be unreliable during the process of visual inspection have been flagged in the second-to-last column of Table \ref{tbl-absorbers}.

As a demonstration of our data quality, the annotated F140W direct image for one of the nine fields (Field 2; SDSS J083842.05+025703.7) is shown in Figure ~\ref{fig:absmatch}, alongside the 2D and 1D G141 grism spectra of the galaxies with impact parameters less than 200 kpc and well-determined redshifts matching to a detected \ion{Mg}{2} absorption system.

\subsubsection{Notes on Individual Fields}
\label{sec:Notes}

\begin{figure*}[!t]
\centering
\begin{center}
\includegraphics[width=1.5\columnwidth]{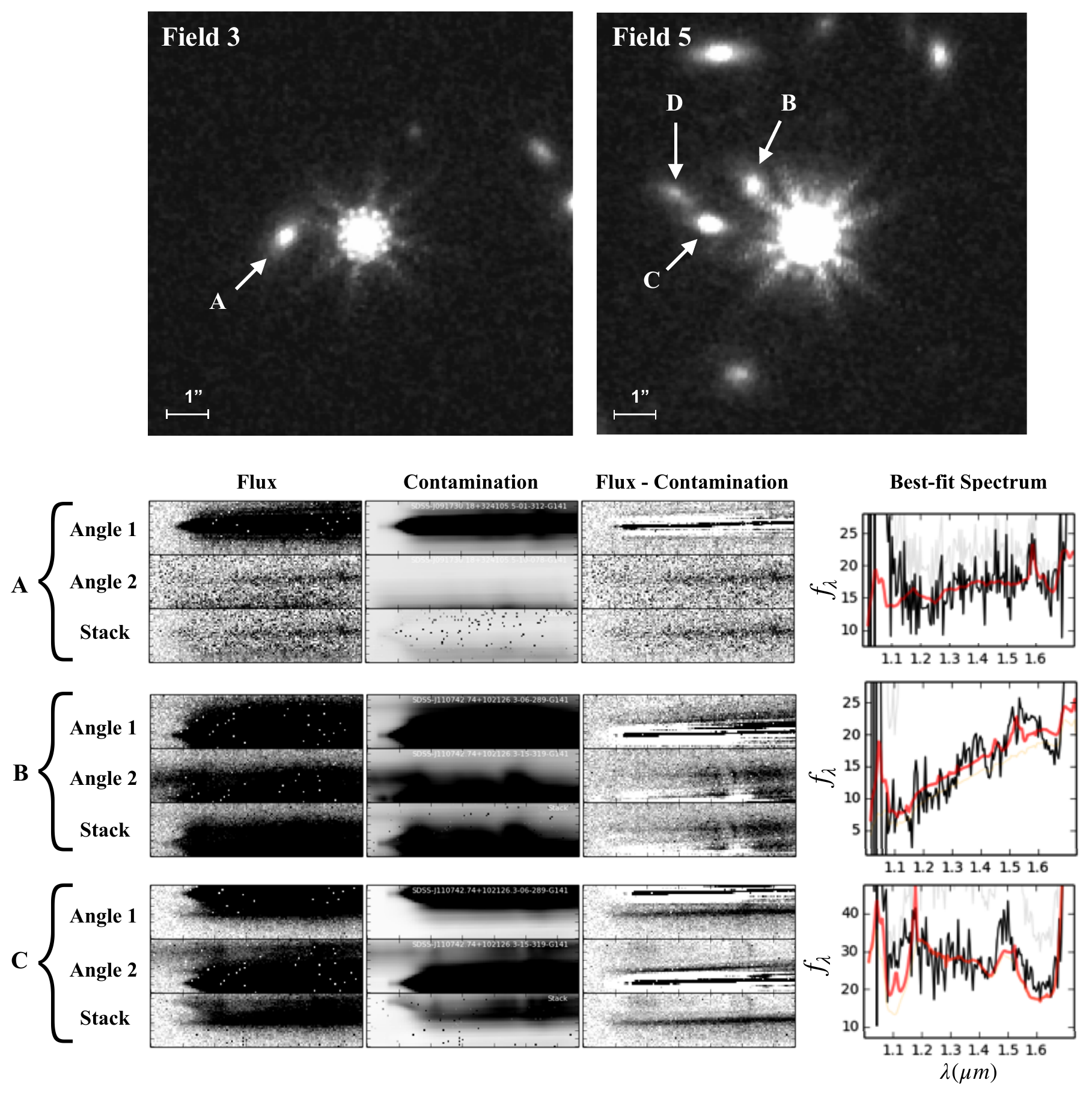}
\caption{\small A $10\arcsec\times10\arcsec$ cutout of the F140W direct image from fields 3 and 5, centered on the quasar in each field.  Shown below are the G141 grism spectra for three of the four galaxies labeled in the images. Galaxies C and D were not separately resolved in the reduced grism data, so the spectrum for galaxy C includes the combined emission from both objects. From left to right we present the 2D raw G141 flux, the modeled G141 contamination from nearby sources, and the contamination-subtracted flux for each of the two orientations and the combined stack. The final column presents the best-fit 1D spectrum for each object with the black trace indicating the observed flux and the red trace indicating the best-fit model spectrum. While galaxy A suffers substantial contamination in one orientation, a clean spectrum and reliable redshift determination has been extracted using the second orientation alone. For contrast, galaxies B and C in field 5 are clearly resolved in the imaging, but their redshifts cannot be determined due to residual contamination in both orientations of the extracted grism spectra. \label{fig:Q-Contam}}
\end{center}
\end{figure*}

Foreground galaxies with small ($<7$ kpc) impact parameters are often difficult to resolve from the quasar PSF, and the orientation of the grism spectra can lead to catastrophic contamination that makes the galaxy redshifts impossible to determine. This means that, for some fields, information about our census of foreground galaxies may be incomplete. As shown in Figure \ref{fig:Q-Contam}, field 5 is problematically crowded, with three galaxies detected within 4$\arcsec$ of the background quasar. While each of these galaxies is clearly resolved in the imaging, galaxy D was not separately resolved from galaxy C in the grism data. Additionally, galaxies B and C each suffered catastrophic contamination from the quasar and another nearby galaxy in both of the randomly chosen orientations for the grism observations. Our inability to extract redshifts for these three galaxies introduced an unacceptable level of uncertainty into our galaxy-absorber matching for this field. Thus, we ultimately excluded it from our full analysis.

\begin{figure*}[!t]
\centering
\begin{center}
\includegraphics[width=1.5\columnwidth]{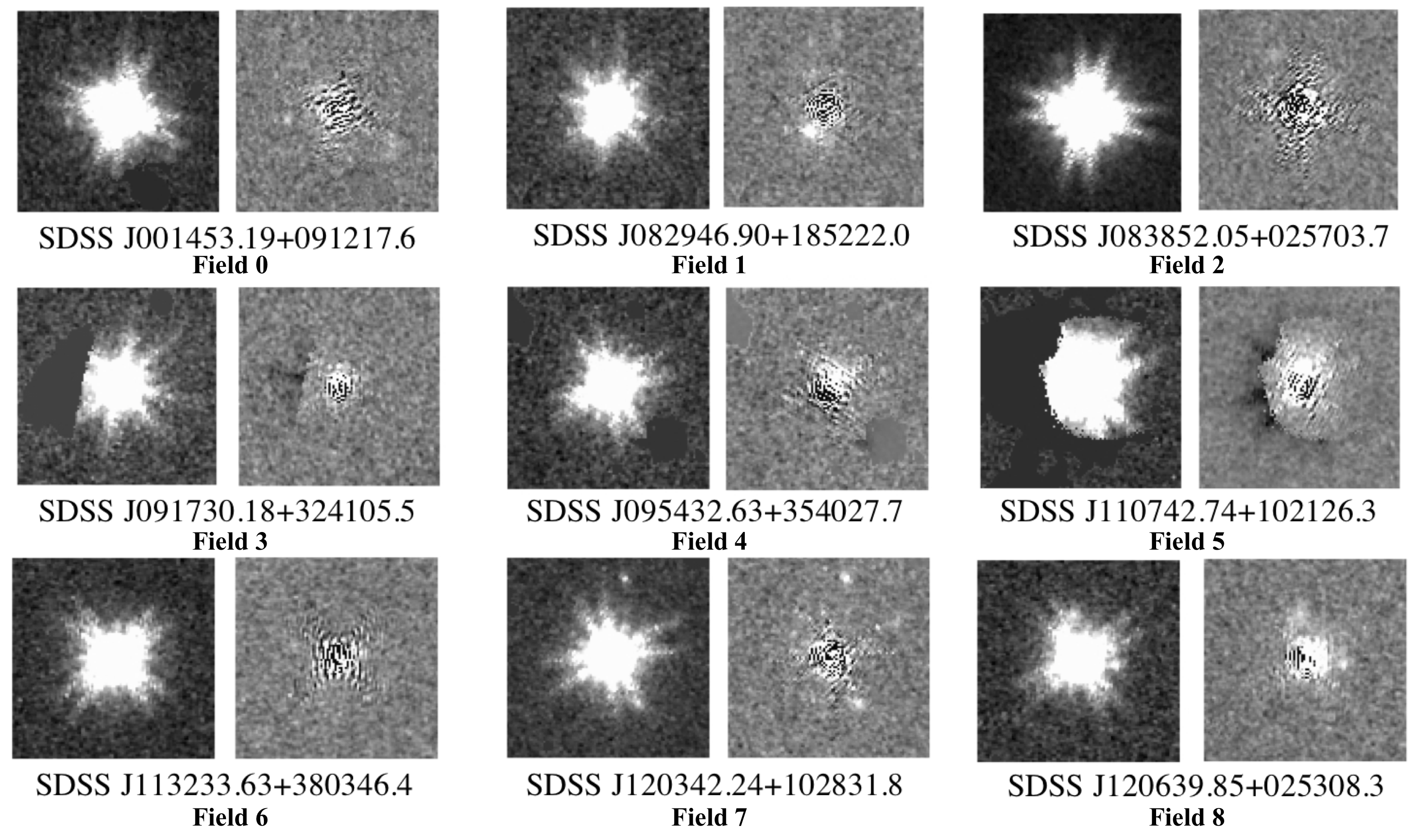}
\caption{\small The quasar from each field before and after PSF subtraction. The scale of each image is $6\arcsec\times6\arcsec$, and resolved neighboring galaxies have been masked out. The residual for quasar SDSS J082946.90+185222.0 (Field 1) reveals a likely galaxy component blended with the quasar PSF. Portions of the quasar PSF in SDSS J091730.18+324105.5 (Field 3) and SDSS J110742.74+102126.3 (Field 5) are also partially obstructed by individually detected foreground galaxies, whose profiles have been masked.  \label{fig:Q-PSF}}
\end{center}
\end{figure*}

For comparison in Figure \ref{fig:Q-Contam} we show that in the case of field 3, which has no crowding, the spectrum for a galaxy with a similarly small impact parameter ($\rho=16$ kpc) could be successfully extracted despite one of the grism orientations being heavily contaminated by the quasar. Fitting the grism data for galaxy A, using the second orientation angle alone, produced a spectrum with a cleanly detected H$\alpha$ emission line, leading to a well-determined redshift of $z=1.415$, which matches a \ion{Mg}{2} absorber at $z_{abs}=1.413$.  Taken together, fields 3 and 5 demonstrate the benefit of multiple grism orientations for obtaining galaxy spectra at small impact parameters to a background quasar and also the limitations of this observational method in crowded fields.

In order to determine whether resolvable galaxies are present at even smaller impact parameters, we used the robust 2D galaxy modeling software package {\it \textit{Galfit}} \citep{Peng2010} to fit and subtract the empirically-determined PSF in each field from the central quasar image. In Figure \ref{fig:Q-PSF}, we present image cutouts of the central quasar from each field with pixels associated with resolved galaxies masked. Beside each cutout is the residual image that resulted from subtracting a scaled PSF from the quasar profile. The residual images in fields 1, 4, 7, and 8, revealed possible unresolved galaxies with small angular separations from quasars SDSS J082946.90+185222.0, SDSS J095432.63+354027.7, SDSS J120342.24+102831.8, and SDSS J120639.85+025308.3, respectively. As we were unable to extract grism spectra for these sources, their redshifts are undetermined. Consequently, these sources are not accounted for in our analysis. 

The quasar PSF subtraction for field 7 reveals two faint sources within 3$\arcsec$ of the quasar, whose small sizes may suggest a high redshift origin. Either of these sources could match the absorption system at $z=1.79606$ observed in the spectrum of SDSS J120342.24+102831.8. This absorber was detected due to the presence of \ion{C}{4} absorption and other lines with similarly high ionization, but it was not included in our analysis due to the apparent absence of associated \ion{Mg}{2} absorption. We discuss the overall completeness of the galaxy-quasar pair sample in more detail in section \ref{Sample_Completeness}.

The PSF-subtracted quasar image in Field 1 (SDSS J082946.90+185222.0) also reveals a compact luminous component in the residual.  This could correspond to the $W_{r}=0.97$\AA~ \ion{Mg}{2} absorber at $z=1.01773$, which is not matched to any other galaxies in our sample, or the $W_{r}=1.52$\AA~ absorber at $z=1.80016$ $-$ the latter of which we ignore due to its redshift being outside of the range used in our analysis. Similarly, the apparent residual component in field 4 could correspond to the $W_{r}=0.31$\AA~ absorber at $z=1.35044$, which remains unmatched to any of the galaxies in our sample. Two other absorbers ($W_{r}=0.43$\AA~ at $z=1.15147$ and $W_{r}=0.92$\AA~  at $z=1.19213$) that are not matched to any galaxies are found in field 6, which has a very flat residual in the PSF-subtracted quasar image and no evidence of excess flux. We can state with high confidence that these two absorption systems are not the result of luminous galaxies at very small impact parameters.

\subsubsection{Galaxy Morphology}

\begin{figure}[!t]
\centering
\includegraphics[width=0.8\columnwidth]{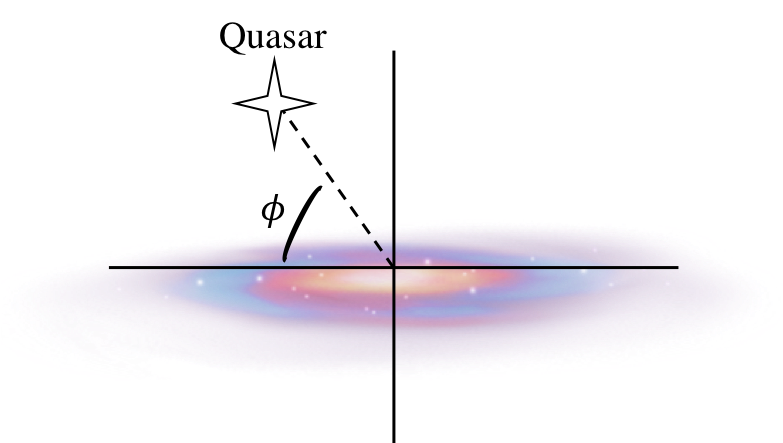}
\caption{\small A diagram illustrating how the azimuthal angle ($\phi$) is defined for the galaxies in this work. \label{fig:AngleFig}}
\end{figure}

The azimuthal angle ($\phi$) of gas relative to the minor axis of a galaxy (see Figure ~\ref{fig:AngleFig}) can be used to infer associations with inflows or outflows around galaxies.  To precisely measure the azimuthal distribution of the \ion{Mg}{2} absorbers around galaxies in our sample requires reliable measurements of the 2D morphologies and orientations of galaxies in the foreground field of each quasar. 
The high resolution and small native pixel scale (0.06\arcsec/pix) of the WFC3 F140W images enable us to examine the morphologies of the galaxies proximate to the quasar sightline.  Using \textit{Galfit}, we have derived the effective radius (R$_{e}$), S{\'e}rsic index, axis ratio (b/a), and position angle for the galaxies within a projected distance of 200 kpc around each quasar. As input to \textit{Galfit}, we produced an empirical point-spread function (PSF) for each field by stacking the normalized profiles of multiple unsaturated stars.  Most fields had several unsaturated stars, and all fields had at least one.  Before fitting, each galaxy image was masked to ignore pixels from neighboring galaxies or stars. The masks were generated using the segmentation maps of each field, which were produced using \textit{SExtractor} \citep{BA96}.

We fit the galaxies with a S{\'e}rsic profile, with initial  parameters for each of the following: integrated magnitude (obtained from \textit{SExtractor}), effective radius, S{\'e}rsic index, axis ratio, and position angle. For cases in which \textit{SExtractor} identified two closely spaced or merging galaxies to be separate sources, we modeled each source individually. In order to obtain a satisfactory fit, some galaxies required a two-component (bulge plus disk) model. In such cases, only the disk component was used to determine the position angle. After \textit{Galfit} converged to the best-fit estimates for these parameters, the model-subtracted residual images were visually inspected to validate the results.  If the \textit{Galfit} modeling diverged or produced catastrophic residuals for all reasonable initial parameters, the morphological measurements were not included in our analysis. We used the best-fit position angle and the image coordinates of the galaxy and the central quasar (obtained from \textit{SExtractor}) to calculate the azimuthal angle with respect to the quasar.

\section{Results}

\subsection{Sample Completeness}
\label{Sample_Completeness}

Eight of the nine fields targeted in this survey were determined to be viable for the analysis of absorber-galaxy correlations.  As we discuss in section \ref{sec:Notes}, the one excluded field (field 5; SDSS J110742.74+102126.3) was affected by the unresolvable blending of multiple galaxies in close angular proximity to the quasar sightline. The data from the remaining eight fields have been used in the analysis described below.

We report an exceptionally high detection rate for high-redshift \ion{Mg}{2} host galaxies, having spectroscopically identified candidate host galaxies for 34/38 (89\%) of the $0.64<z<1.6$ \ion{Mg}{2} absorbers with confident identifications in the eight viable fields.  This rate of detection is equivalent to that reported by \citet{Hamanowicz20}, who used MUSE integral field unit (IFU) observations to detect the [O II] emission from hosts of 14 HI-selected \ion{Mg}{2} absorbers at $z<1$.  

Another recently published study by \citet{Schroetter19} utilized the MUSE IFU and UVES spectrograph at the ESO Very Large Telescope to obtain a 75\% detection rate for a sample of 79 \ion{Mg}{2} absorbers selected with a similar range in redshift and equivalent widths to our sample.  Using [OII] emission to estimate the star formation rates of galaxies in their 22 quasar fields, \citet{Schroetter19} reached a similar sensitivity to our H$\alpha$ observations.  The slightly higher detection rate that we achieve could be due in part to the lower spectral resolution of the G141 grism data, which required us to adopt a wider threshold for the redshift matching criterion ($|\Delta z|/(1+z)<0.006$) we used to match galaxies with absorbers, while the VLT redshifts of \citet{Schroetter19} achieve a resolution of a few km s$^{-1}$ and can thus be matched or ruled out as hosts of the target \ion{Mg}{2} absorption with higher precision.

\begin{figure}[t!]
\includegraphics[width=1.0\columnwidth]{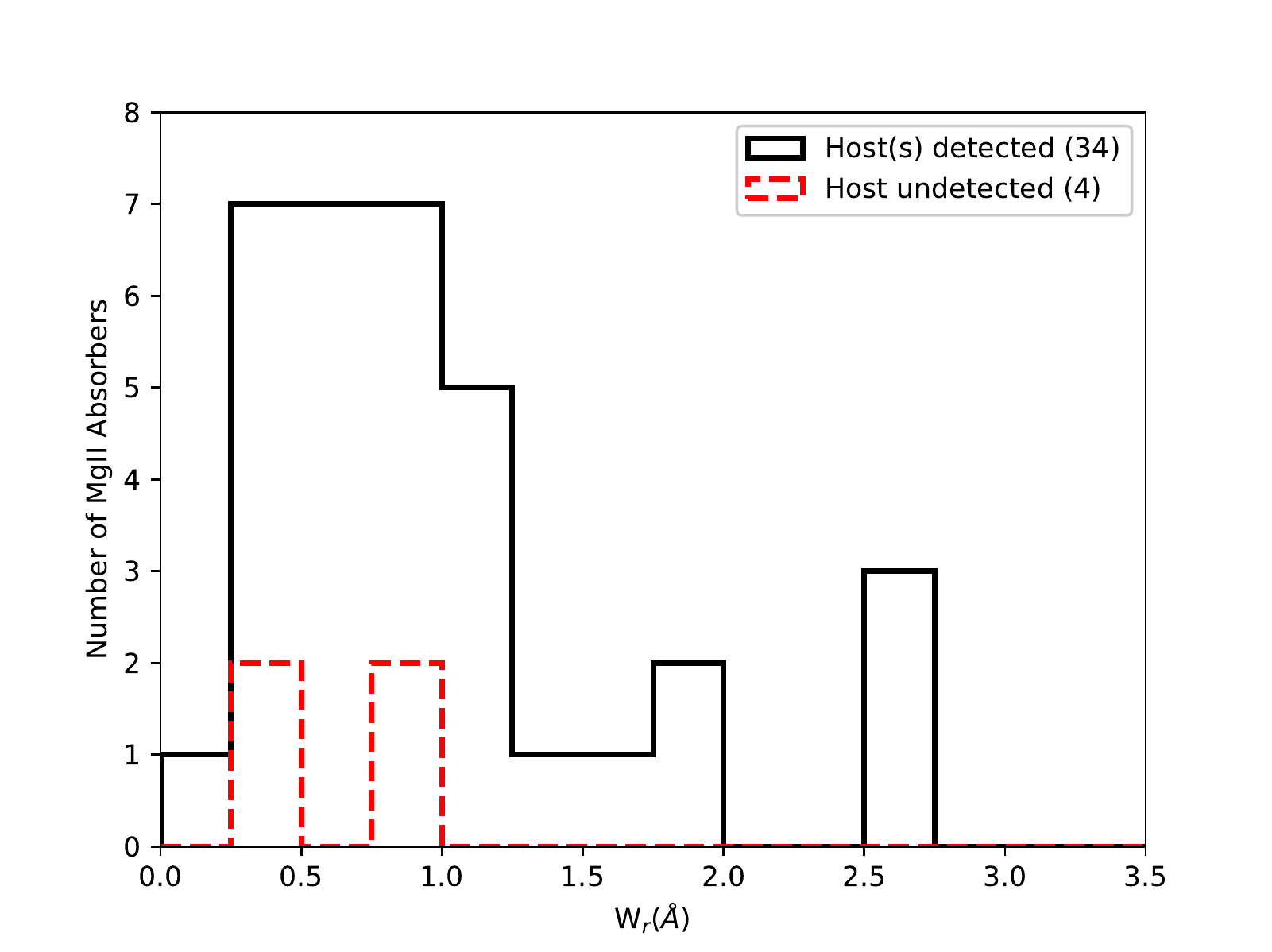}
\caption{\small The absorber rest-frame equivalent width ($W_{r}$) distribution for 2796$\AA$ \ion{Mg}{2} absorbers with and without at least one spectroscopically confirmed galaxy counterpart. This figure includes only absorbers in the redshift range where H$\alpha$ emission is detectable in the G141 grism data ($0.64<z<1.6$) and excludes absorbers in field 5. \label{fig:Wdist_found_unfound}}
\end{figure}

Any \ion{Mg}{2} host galaxies that remain unaccounted for in our study could be fainter than the H$\alpha$ flux limit of our sample, highly dust-obscured, or have too small an angular separation from the quasar so as to be resolvable. In a closer analysis of the quasar PSFs, described in the previous section, we have tested the latter possibility.  Four absorption systems that lack detected galaxy counterparts are observed in Fields 1, 4, and 6 -- the first two of which have notable residual features in the PSF-subtracted quasar images, which could indicate at least one unresolved galaxy at a small impact parameter.  For reasons described in detail in the next section, one might expect galaxies at small impact parameters to produce absorption systems with high equivalent widths \citep[e.g.,][]{B12a}. However, as shown in Figure \ref{fig:Wdist_found_unfound}, the equivalent width distribution of the four absorbers that lack a spectroscopically confirmed galaxy match in our sample is not skewed toward high equivalent widths.  This result suggests that the galaxies blended with the quasar PSF are not the missing hosts of the four absorbers lacking galaxy matches. However, we also cannot rule out the possibility that some of higher equivalent width \ion{Mg}{2} absorbers in our sample that have spectroscopically confirmed galaxy matches may also match to one or more of these galaxies at very small impact parameters, whose redshifts remain undetermined.

\subsection{$W_{r}^{2796}$ $-$ Impact Parameter Relation}
\label{sec:EW_b}

\begin{figure}[t!]
\includegraphics[width=1.0\columnwidth]{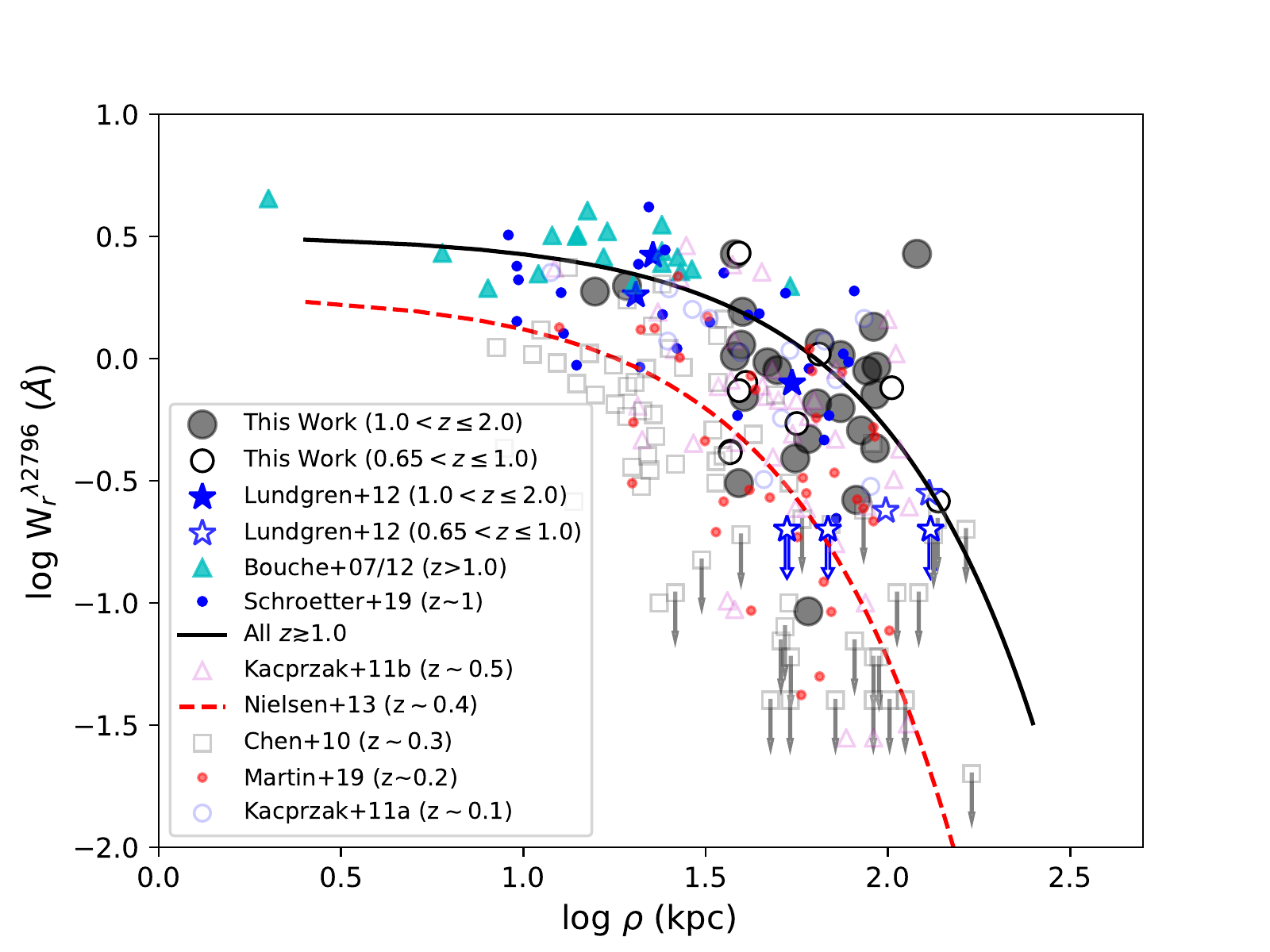}
\caption{\small The \ion{Mg}{2} 2796\AA~ absorber rest-frame equivalent width ($W_{r}$) plotted versus impact parameter ($\rho$) for galaxies matched to \ion{Mg}{2} absorption in this work, shown in comparison to other surveys in the literature. A log-linear fit to the largest samples at intermediate ($z\sim0.4$) \citep{Nielsen2013} and higher redshift ($z\sim1.2$; \citet{L12}, \citet{B12b}, \citet{Schroetter19}, and this work) is overplotted, indicating possible evolution of the gas radius of \ion{Mg}{2} absorbing galaxies. \label{fig:WvsRho}}
\end{figure}

Samples of spectroscopically-confirmed galaxy-absorber pairs have gradually grown over the past few decades, revealing a now well-established anti-correlation between galaxy impact parameter and the associated equivalent width of \ion{Mg}{2} \citep[e.g.,][]{LB90, BB91, B06, Kacprzak08, Chen10a, Churchill13}. This relation has been suggested to explain the strong observed correlation between \ion{Mg}{2} equivalent width and dust extinction in background quasars \citep{Menard08}, as well as the strong correlation between \ion{Mg}{2} equivalent width and [\ion{O}{2}] emission observed in stacks of SDSS absorption spectra \citep{Menard11}.  In Figure \ref{fig:WvsRho} we plot the \ion{Mg}{2} absorber rest-frame equivalent width versus galaxy impact parameter for high-confidence absorber-galaxy pairs. We find agreement with previous results that indicate an inverse correlation with $\sim1$ dex of scatter.

In Figure ~\ref{fig:WvsRho} we have overplotted log-linear fits of the form $\log(W_{r}) = a*\rho+b$ applied by \citet{Nielsen2013} to a large sample with $z\sim0.4$ (where $a = -0.015 \pm 0.002$ and $b = 0.27 \pm 0.11$). Fitting the same form to a combined sample of $z>1$ pairs from this work, together with the previous works of \citet{B12a}, \citet{L12}, and \citet{Schroetter19}, gives best-fit coefficients of $a = -0.008 \pm 0.001$ and $b = 0.51 \pm 0.03$.  The overlaid results, shown in Figure \ref{fig:WvsRho}, suggest a significant evolution in the relation from $z\sim1.5$ to $z\sim0.4$, consistent with circumgalactic \ion{Mg}{2} gas extending to larger physical radii around galaxies at earlier times. This could potentially be explained by galaxy-scale winds extending to greater distances at an epoch when galaxies typically had lower masses and higher star formation rates. Our findings appear to agree with \citet{Lan2020}, which recently reported evolution in the covering fraction of strong \ion{Mg}{2} absorption around star-forming galaxies in this same redshift range. However, we caution that the inhomogeneous selection of targets in the included data sets in Figure ~\ref{fig:WvsRho} could also potentially complicate the redshift evolution implied by this comparison.

In a recent study of 27 \ion{Mg}{2} absorbers detected around 228 galaxies at $0.8<z<1.5$, \citet{Dutta2020} reported an absence of evolution in the $W_{r}-\rho$ relation compared to literature measurements at $z\sim0.5$. This discrepancy with our finding could potentially be explained by the fact that the \citet{Dutta2020} sample was selected blind to the incidence of \ion{Mg}{2} in the spectra of the targeted quasars. In contrast, our $z\sim1.2$ sample combines surveys of absorption-selected galaxies, which could be biased to tracing outflows and not represent the isotropic average of the radial profile of \ion{Mg}{2} around an unbiased sample of galaxies. It is also important to note that the lower redshift ($z\sim0.5$) measurements from \citet{Nielsen2013}, to which both we and \citet{Dutta2020} have compared our higher redshift measurements of the $W_{r}-\rho$ relation, were obtained through a heterogeneous selection that also includes \ion{Mg}{2}-selected quasar fields. 

With the possible exception of galaxies with very low cold gas covering fractions (e.g., luminous red galaxies; \citet{Bowen11}, but see also \citet{Zhu2014, PR2015}), this anti-correlation appears to hold for a diverse range of galaxy types and can be physically interpreted, to first-order, as a decreasing covering fraction of low-ionization metal-enriched gas with increasing distance from an average galaxy.  The persistent and significant scatter observed in the W$_{r}-\rho$ relation may be caused by multiple complicating and compounding factors, including: galaxy inclination and azimuthal angle (i.e., the angular position of the detected gas relative to the major axis of the associated galaxy), stellar mass, star formation rate, and environment.

\citet{Bordoloi11} and \citet{Hamanowicz20} have observed that the $W_{r}-\rho$ relation for absorbers matched to multiple galaxies is offset compared to individual galaxies, strengthening the case that environment affects the column density and velocity dispersion of a detected absorption system. Some studies \citep[e.g.,][]{Chen10b, Kacprzak11b, Nielsen2013} have also reported success at reducing the scatter in this relation by accounting for galaxy luminosity, inclination, and mass. In Section \ref{sec:modeling} we describe a new model for predicting the W$_{r}$ of a system, by accounting simultaneously for the azimuthal angle, impact parameter, and environment of galaxy-absorber pairs.

\begin{figure}[t!]
\includegraphics[width=1.0\columnwidth]{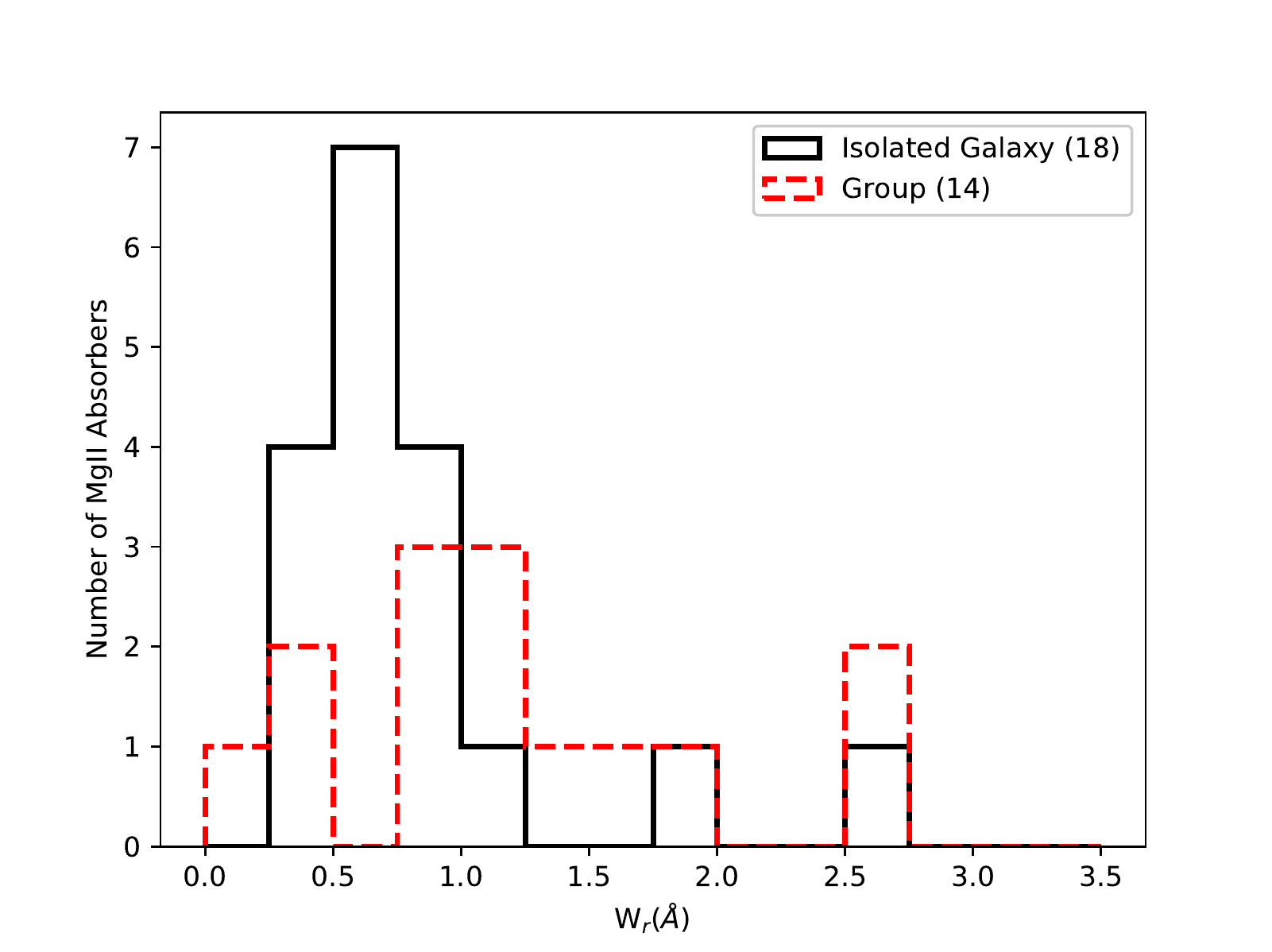}
\caption{\small The absorber rest frame equivalent width distribution of $W_{r}^{\lambda2796}$ for \ion{Mg}{2} absorbers in this sample, as a function of galaxy environment. The mean equivalent width of absorbers matched to multiple galaxies in a field is found to be marginally higher compared to absorbers matched to only one galaxy. A K-S test returns a $96\%$ probability that the two populations of absorbers trace different underlying distributions in equivalent width. \label{fig:Wdists_env}}
\end{figure}

\subsection{\ion{Mg}{2} Absorber Environments}

Recent IFU studies with improved methods for detecting galaxies in quasar foregrounds have found that it may be common for multiple massive galaxies to be matched to a single absorption system \citep[e.g.,][]{Bielby17, Peroux17, Klitsch18, Rahmani18, Peroux19, Hamanowicz20, Bielby2020}. Since our data enable redshift estimates for all galaxies with SFR $\gtrsim 1$ M$_{\odot}$ yr$^{-1}$ in each quasar field, we have also been able to investigate the typical environments of \ion{Mg}{2} absorbers at $z\sim1.2$.

As shown in Figure~\ref{fig:Wdists_env}, nearly half (14/32) of the \ion{Mg}{2} absorbers with H$\alpha$ coverage in our survey are spectroscopically matched to more than one galaxy within 200 kpc of the quasar sightline.  Furthermore, we find that the \ion{Mg}{2} absorption equivalent width ($W_{r}$) exhibits a weak correlation with environment. The mean \ion{Mg}{2} absorption equivalent width for systems matched to apparently isolated galaxies is $0.84\pm0.14$\AA, compared to a mean of $1.21\pm0.20$\AA~ for absorbers matched to multiple galaxies.

These findings appear to agree with the recent results of \citet{Dutta2020} and \citet{Fossati2019}, whose observations of the MUSE Ultra Deep Field revealed enhanced absorption of \ion{Mg}{2} around galaxies in groups, compared to galaxies in more isolated environments.  These results suggest that gas stripping from gravitational interactions between galaxies in these dense environments increases the cross-section of cool gas and lead to an increased equivalent width in the associated \ion{Mg}{2} absorption.  Evidence for multiple galaxies contributing to intra-group gas has also been noted by other recent studies \citep{Whiting06, Kacprzak10, Gauthier13, Johnson2015}.

\subsection{\ion{Mg}{2} Incidence and Star Formation Rate}

\begin{figure}[!t]
\includegraphics[width=1.0\columnwidth]{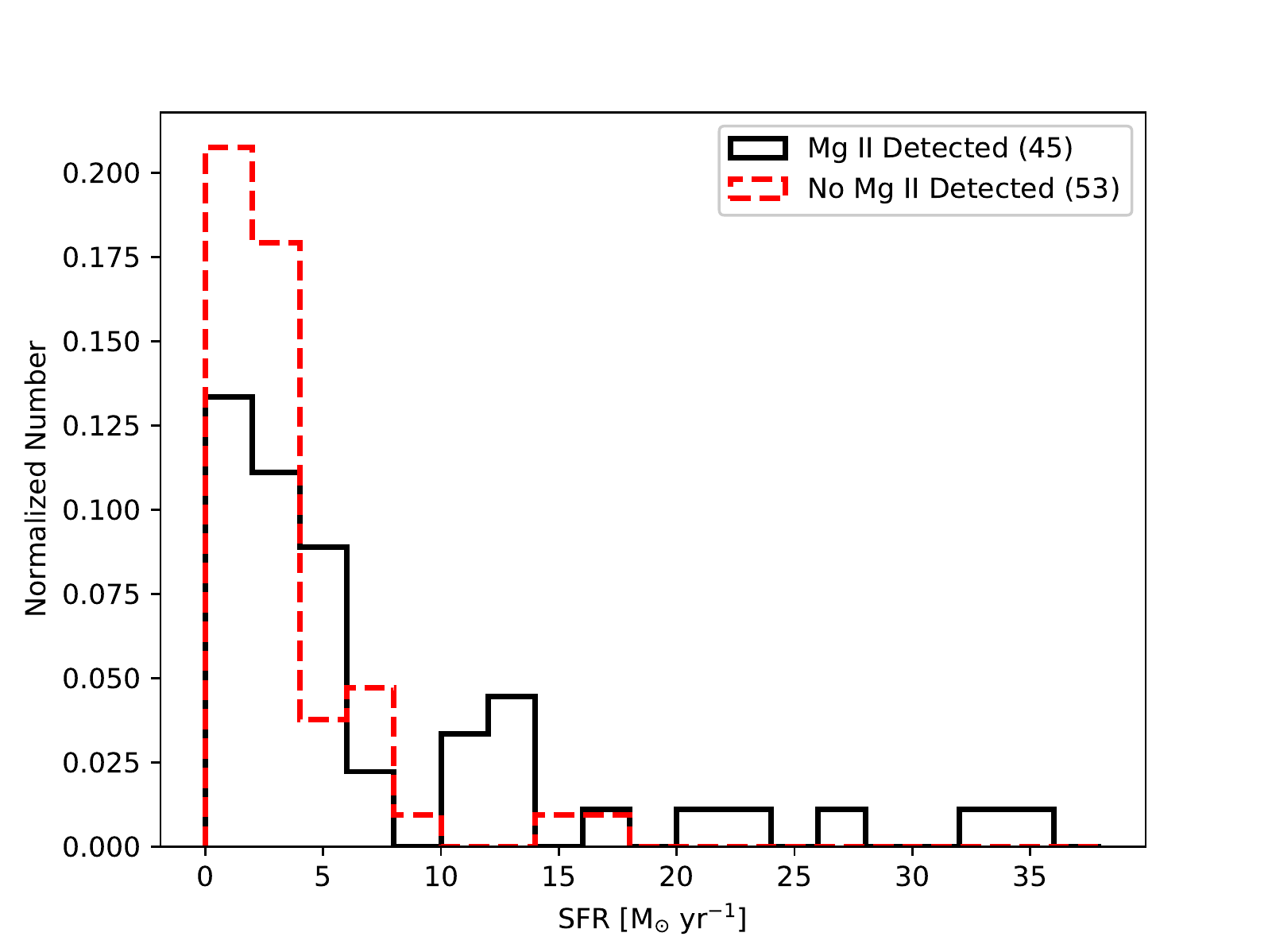}
\caption{\small The star formation rate distribution for galaxies with H$\alpha$ emission measurements above our detection threshold and within 150 kpc of the quasar sightline. Galaxies with coincident \ion{Mg}{2} absorption in the quasar sightline have significantly higher average SFRs and trace a significantly different distribution of SFRs compared to galaxies without \ion{Mg}{2} detections, as quantified with $>99$\% probability by a K-S test.\label{fig:SFRdist_found_unfound_150}}
\end{figure}

An abundance of evidence now supports a correlation between the star formation rate and the covering fraction of circumgalactic \ion{Mg}{2} absorption in galaxies \citep[e.g.,][]{Zibetti07, Menard08, NSM10, Menard11, Bordoloi11, Menard12, Bordoloi14b, Lan2018}.  Analyses of the largest samples of \ion{Mg}{2} compiled to date have even suggested that evolution of the redshift number density of strong \ion{Mg}{2} absorbers can be used to trace the cosmic star formation history from $z\sim6$ \citep{MS12, ZM13}.  

 While we have provided in Table \ref{tbl-absorbers} a full accounting of absorption-matched galaxies detected out to impact parameters of 200 kpc, we will hereafter truncate our analysis at $\rho<150$ kpc in order to minimize the uncertainties introduced by including galaxies with very large physical separations, which are statistically less likely to be truly associated with the detected absorption. In an examination of all galaxies with SFR measurements above our detection threshold and within 150 kpc of the quasar sightline, we find that galaxies associated with \ion{Mg}{2} absorption have significantly higher average SFRs compared to galaxies for which no \ion{Mg}{2} was detected in the quasar spectrum (see Figure ~\ref{fig:SFRdist_found_unfound_150}). The mean SFR for galaxies with detected \ion{Mg}{2} is $7.2\pm1.3$ M$_{\odot}$ yr$^{-1}$, compared to $2.9\pm0.5$~M$_{\odot}$ yr$^{-1}$ for galaxies without a \ion{Mg}{2} detection.  

In the local universe, galaxy-scale outflows of photo-ionized $T\sim10^{4}~K$ gas have been observed in association with galaxies with higher than average star formation rate surface densities \citep{Chisholm15,Heckman15,Heckman_Borthakur16,Ho2016,Chisholm17}. These findings support theoretical models in which not only the rate of star formation, but also its physical concentration, determines a galaxy's capacity for launching large-scale outflows. Supporting this model, a recent analysis by \citet{Bordoloi14b} reported a strong correlation between the equivalent width of outflowing \ion{Mg}{2} gas and the star formation rate surface density ($\Sigma_{\rm{SFR}}$) in 486 zCOSMOS galaxies at $1 <z <1.5$. If a substantial fraction of the \ion{Mg}{2} detected in the extended halos of galaxies in our sample originated in star formation-driven winds (rather than, e.g., tidal stripping or infall from the intergalactic medium), we might also expect to observe a correlation between $\Sigma_{\rm{SFR}}$ and the incidence of circumgalactic \ion{Mg}{2} absorption.

\begin{figure}[!t]
\includegraphics[width=1.0\columnwidth]{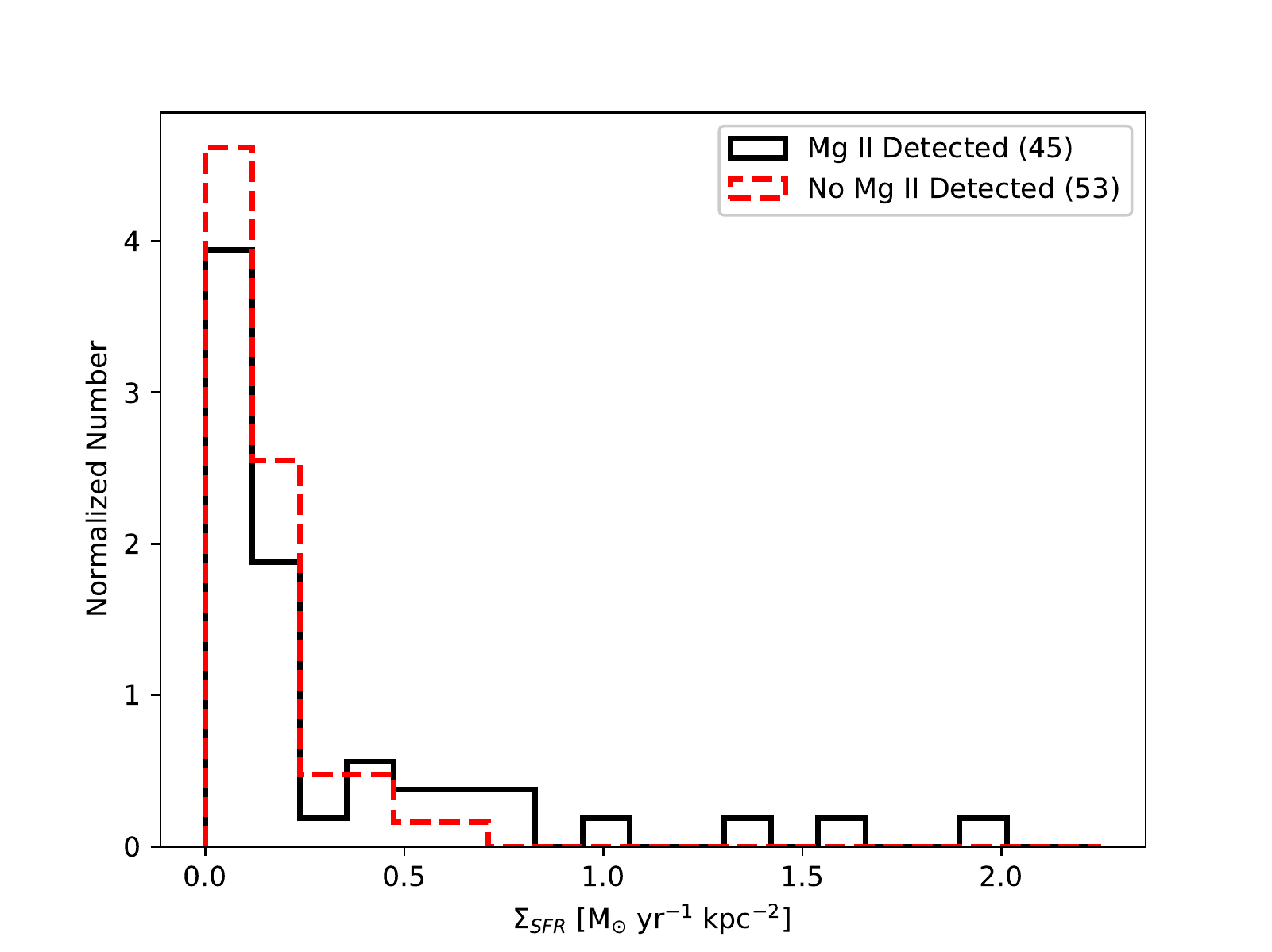}
\caption{\small The star formation rate surface density distribution for galaxies with SFR measurements above our detection threshold and within 150 kpc of the quasar sightline. Galaxies matched to \ion{Mg}{2} absorption in the quasar sightline have higher average  $\Sigma_{\rm{SFR}}$ compared to those without detected absorption, but the difference in the distributions for the two populations is not highly significant (as quantified by a K-S statistic of 0.22, with a p-value of 0.15).\label{fig:SFRD}}
\end{figure}

In order to estimate the $\Sigma_{\rm{SFR}}$ for galaxies in our sample, we produced spatially resolved maps of the H$\alpha$ emission following the methodology described in \citet{L12}. However, for most individual galaxies the signal-to-noise ratio was too low to precisely measure the morphology of the H$\alpha$ emitting region. Thus, we have estimated the galaxy sizes using the effective radii extracted from the F140W imaging, which captures the rest-frame stellar continuum emission of the galaxies at the redshift range of our study.  We calculate the average star formation rate surface density, $\Sigma_{\rm{SFR}}$, as:
\begin{equation}\label{eq:SFRSD}
\Sigma_{\textrm{SFR}}[\textrm{M}_{\odot}~\textrm{yr}^{-1}~\textrm{kpc}^{-2}] = \frac{\dot{M_{\odot}}}{\pi R_{e}^{2}}
\end{equation}
where $\dot{M_{\odot}}$ is the H$\alpha$-derived SFR (Equation ~\ref{eq:SFR}) and R$_{e}$ is the effective radius enclosing 50\% of the flux in the F140W continuum emission, measured in kpc.

In an analysis of H$\alpha$ emission line maps from 57 typical star-forming galaxies at $z\sim1$, \citet{Nelson12} reported that H$\alpha$ emission broadly follows the profile of the stellar continuum but is slightly more extended, such that $R_{e,H\alpha}/R_{e,F140W}\sim1.3$ \citep{Nelson12}, consistent with the expectations from models of inside-out growth in disk galaxies (see also \citet{Nelson2016}). While measuring galaxy structural parameters from the stellar continuum emission may slightly underestimate the effective radius enclosing active star formation, the assumption of a circularized profile could also lead to an overestimation of the true star-forming area of the galaxies, since star formation tends to be concentrated in clumps within galaxy disks \citep{Rubin2010, Kornei2012}. Acknowledging these opposing uncertainties, we determined that the benefit of higher signal to noise and F140W imaging was sufficient to justify its use in estimating the area of active star formation.

As shown in Figure ~\ref{fig:SFRD}, the distribution in $\Sigma_{\rm{SFR}}$ for our sample of galaxies extends to $\sim2$~M$_{\odot}$yr$^{-1}$kpc$^{-2}$, in agreement with similar measurements for typical galaxies in this same redshift range \citep[e.g.,][]{Nelson2016}. We observe a  marginally significant difference between the $\Sigma_{\rm{SFR}}$ distributions for galaxies with and without detected \ion{Mg}{2} absorption.  The mean $\Sigma_{\rm{SFR}}$ for galaxies matched to a \ion{Mg}{2} detection is $0.30\pm0.06$~M$_{\odot}$yr$^{-1}$ kpc$^{-2}$, compared to $0.14\pm0.02$~M$_{\odot}$yr$^{-1}$ kpc$^{-2}$ for galaxies without a \ion{Mg}{2} detection. However, we note that the median of the $\Sigma_{\rm{SFR}}$ distribution is not statistically offset between the galaxies with a \ion{Mg}{2} detection ($0.12\pm0.08$ yr$^{-1}$ kpc$^{-2}$) and those without ($0.10\pm0.02$ yr$^{-1}$ kpc$^{-2}$). A K-S test confirms that the difference in the $\Sigma_{\rm{SFR}}$ distributions for the two populations of galaxies -- matched and unmatched to \ion{Mg}{2} -- is less statistically significant, having only 85\% confidence, compared to our findings for the SFR distributions for the same populations, which indicated a statistically significant difference with 99\% confidence.  

\begin{figure}[t!]
    \includegraphics[width=1.0\columnwidth]{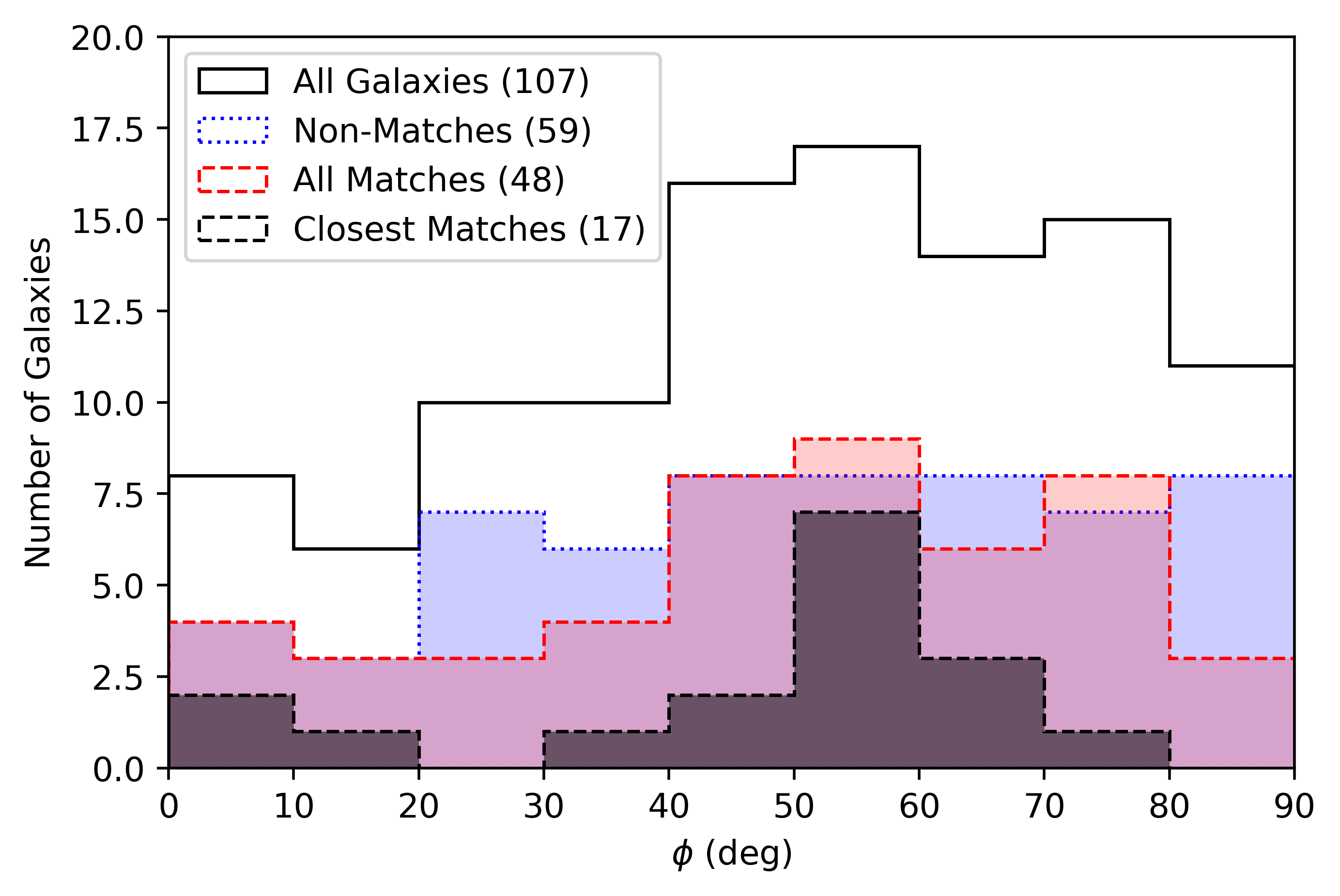}
    \caption{\small The azimuthal angle ($\phi$) distribution of background quasar sightlines probing foreground galaxies in our sample. 
    The plot includes galaxies not matched to \ion{Mg}{2} absorption (blue), all cases where a galaxy matches to a \ion{Mg}{2} detection (red), and a subset of the latter that includes only the matched galaxies with the closest projected separation of all possible pairs (black, filled).}
    \label{fig:AnzimuthalHist}
\end{figure}

The less pronounced difference we observe in the $\Sigma_{\rm{SFR}}$ distributions relative to the SFR distributions for galaxies with and without \ion{Mg}{2} could be due to the additional uncertainty in the measurement of $\Sigma_{\rm{SFR}}$, which is introduced when estimating the area of active star formation in the galaxies. We also note that there is a strong empirical correlation between dust content and SFR, stemming from the fact that galaxies with higher SFRs tend to be more massive galaxies, which are more metal-rich.  Thus, the fact that our SFR measurements are not dust-corrected should lead to a bias in the estimated galaxy SFRs, such that galaxies with higher SFRs will have their rates of star formation more greatly underestimated.  This bias effectively reduces the dynamic range of our estimated SFRs and should be expected to flatten any trends we observe with SFR. Future observations with deeper broadband imaging or G141 grism data could be used to improve the estimates of the size of the star-forming regions in these galaxies and help to better resolve any underlying physical differences, if they exist, in the $\Sigma_{\rm{SFR}}$ of galaxies with and without circumgalactic \ion{Mg}{2} absorption.

An examination of the redshift distributions of the galaxies with and without \ion{Mg}{2} absorption indicates that the galaxies matched to \ion{Mg}{2} skew very slightly toward higher redshift.  The mean (median) redshift of galaxies matched to \ion{Mg}{2} is $1.19\pm0.04$ ($1.23\pm0.05$), compared to $1.08\pm0.04$ ($1.14\pm0.05$) for the galaxies unmatched to \ion{Mg}{2}. Given that \ion{Mg}{2} detections are associated with galaxies that have higher average SFRs, one might expect a Malmquist bias in a flux-limited sample such as ours, favoring the detection of \ion{Mg}{2}-absorbing galaxies at higher redshifts.  However, all of the SFRs estimated for all galaxies in both the matched and unmatched samples lie above the  stated detection threshold. A physical mechanism could instead explain this small redshift difference, since the redshift number density of strong \ion{Mg}{2} absorption lines steeply increases with redshift from $z=0$ to $z\sim2$ \citep{Nestor05, ZM13}, as does the average SFR of galaxies \citep[e.g.,][]{HopkinsBeacom2006}.

\subsection{Azimuthal Distribution of Mg II Absorption}\label{sec:azang}

Outflows containing $T\sim10^{4}~K$ gas appear to be a common feature of star-forming galaxies \citep{Heckman1990, Heckman2000, Steidel1996, Franx1997, Pettini2000, Pettini2001, Shapley2003, Martin2005, Rupke2005, Tremonti2007, Weiner2009, Erb2012}.  In the local universe large-scale outflows have been observed to propagate parallel to the minor axis of galaxies \citep[e.g.,][]{Strickland04, Heckman1990, LH96, Cecil01, SH09}. If the cold gas is preserved out to large radii, the signatures of these winds are expected to be observable statistically in the azimuthal distribution of circumgalactic gas.  The preferential distribution of strong \ion{Mg}{2} along the minor axis of star-forming galaxies was first reported in the stacked absorption line profiles of intermediate redshift galaxies in zCOSMOS \citep{Bordoloi11}.  \citet{Kacprzak12} confirmed a bimodal azimuthal distribution of \ion{Mg}{2} absorption around $z\sim0.5$ host galaxies, finding that the bulk of these absorbers were detected within 20$^{\circ}$ of the major or minor axis of the host galaxy.  \citet{B12b} reported an even more striking bimodality in galaxies at $z\sim0.1$, although, some of this difference could be attributed to the more restrictive selection of inclined disk-dominated galaxies and a different range of W$_{r}$ for \ion{Mg}{2}-absorbing galaxies included in the \citet{B12b} analysis.  

\begin{figure}[t!]
    \includegraphics[width=1.0\columnwidth]{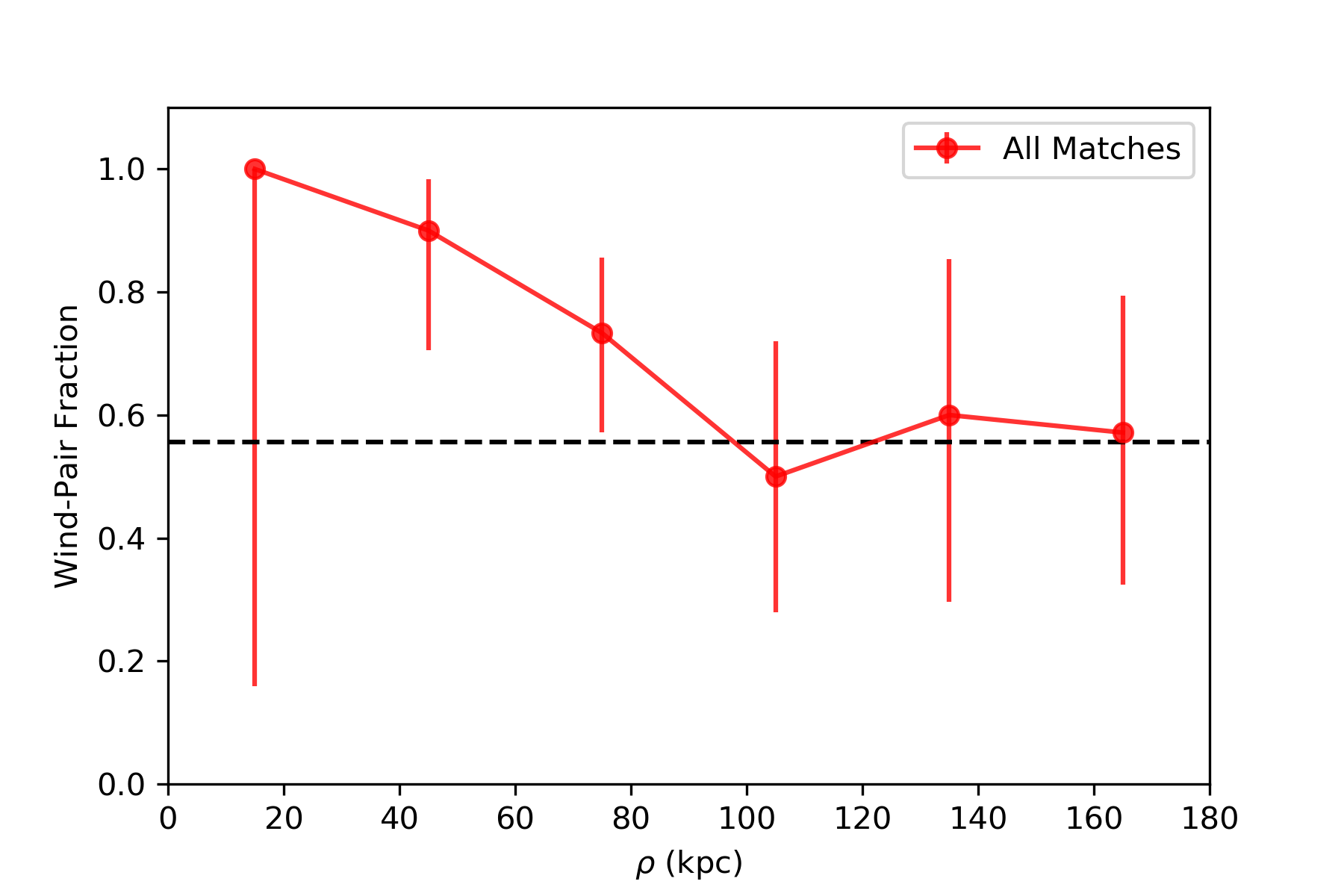}
    \caption{\small For galaxies with a matching \ion{Mg}{2} detection, the fraction with $\phi>40\degree$ (i.e., "wind-pairs", located within 50$\degree$ of the galaxy's semi-minor axis) is plotted as a function of impact parameter. The dashed line indicates the expected fraction for the case of an isotropic distribution.}
    \label{fig:fracwind}
\end{figure}

\begin{figure*}
    \centering
    \includegraphics[width=0.45\textwidth]{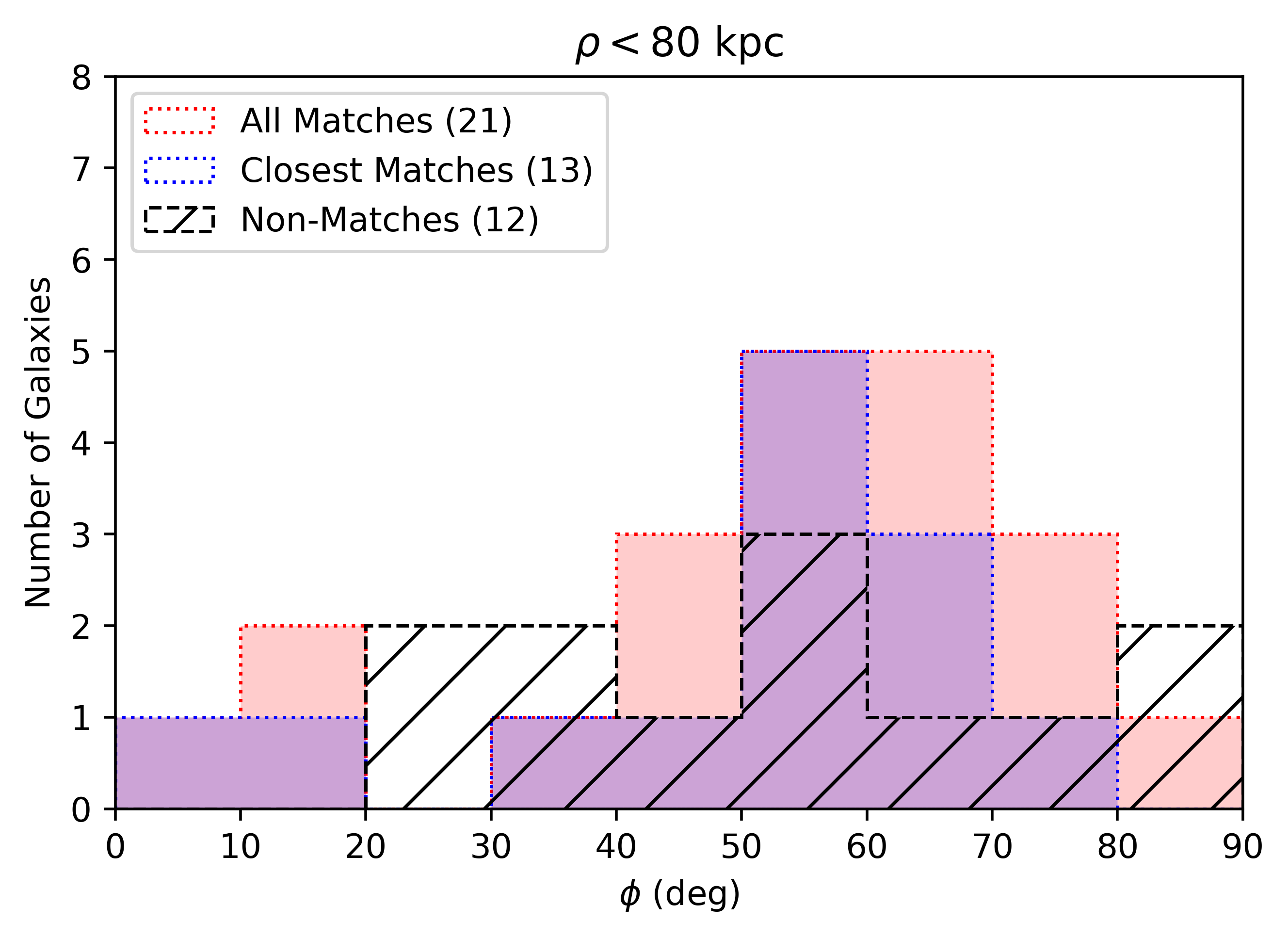}
    \includegraphics[width=0.45\textwidth]{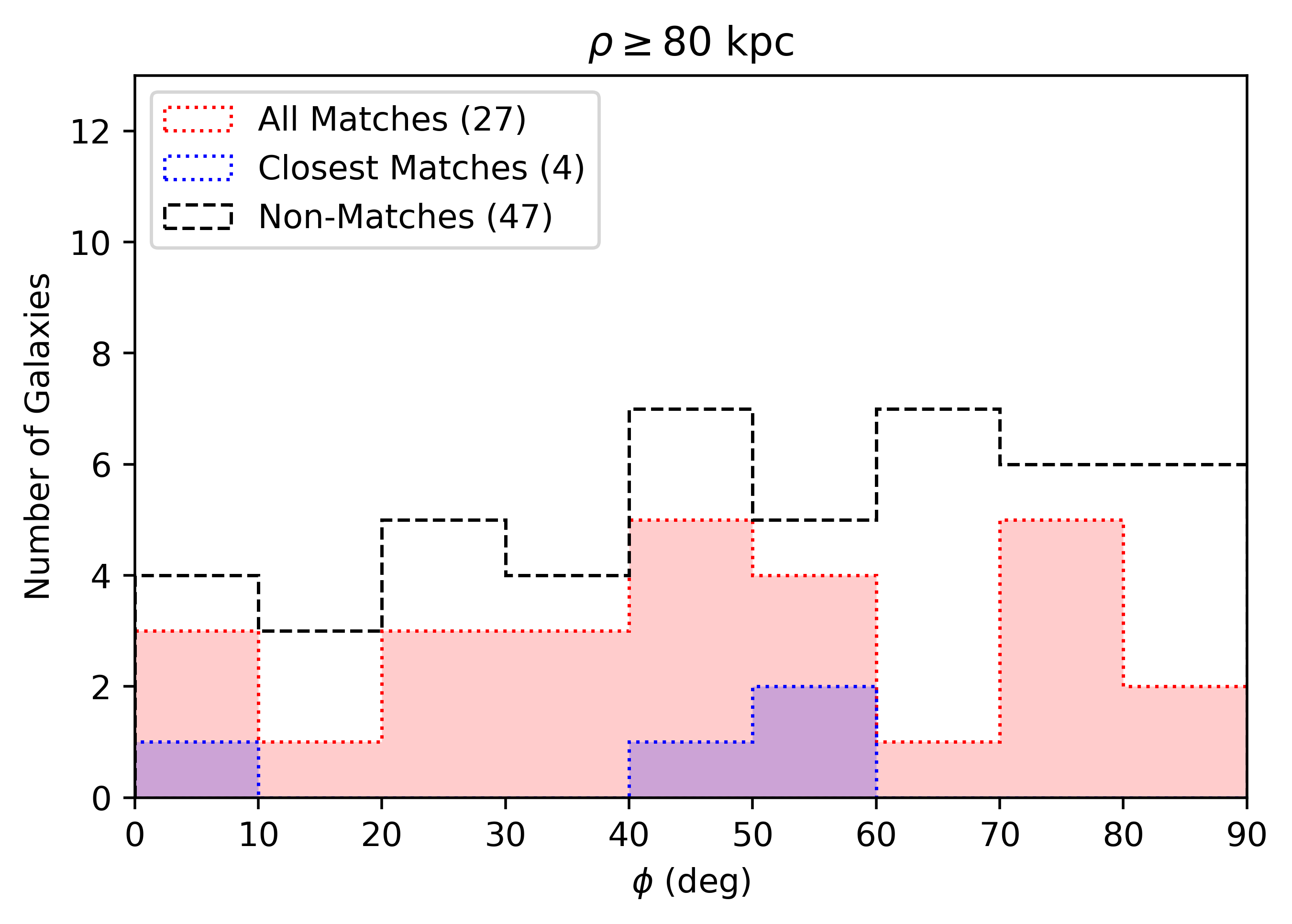}
    \caption{\small The  dependence of the \ion{Mg}{2} azimuthal angle distribution on impact parameter. Galaxies with $\rho < 80$~kpc are shown on the left, and galaxies with $\rho \geq 80$~kpc are shown on the right. The plot includes galaxies not matched with \ion{Mg}{2} absorption (black), closest matches (blue), and all matches (red).  Closest matches are defined as those galaxies that have the smallest impact parameter of all possible  matches to \ion{Mg}{2} absorption in the field where they are detected.
    }
    \label{fig:AnzimuthalHist_BCut}
\end{figure*}

\begin{figure*}
    \centering
    \includegraphics[width=0.45\textwidth]{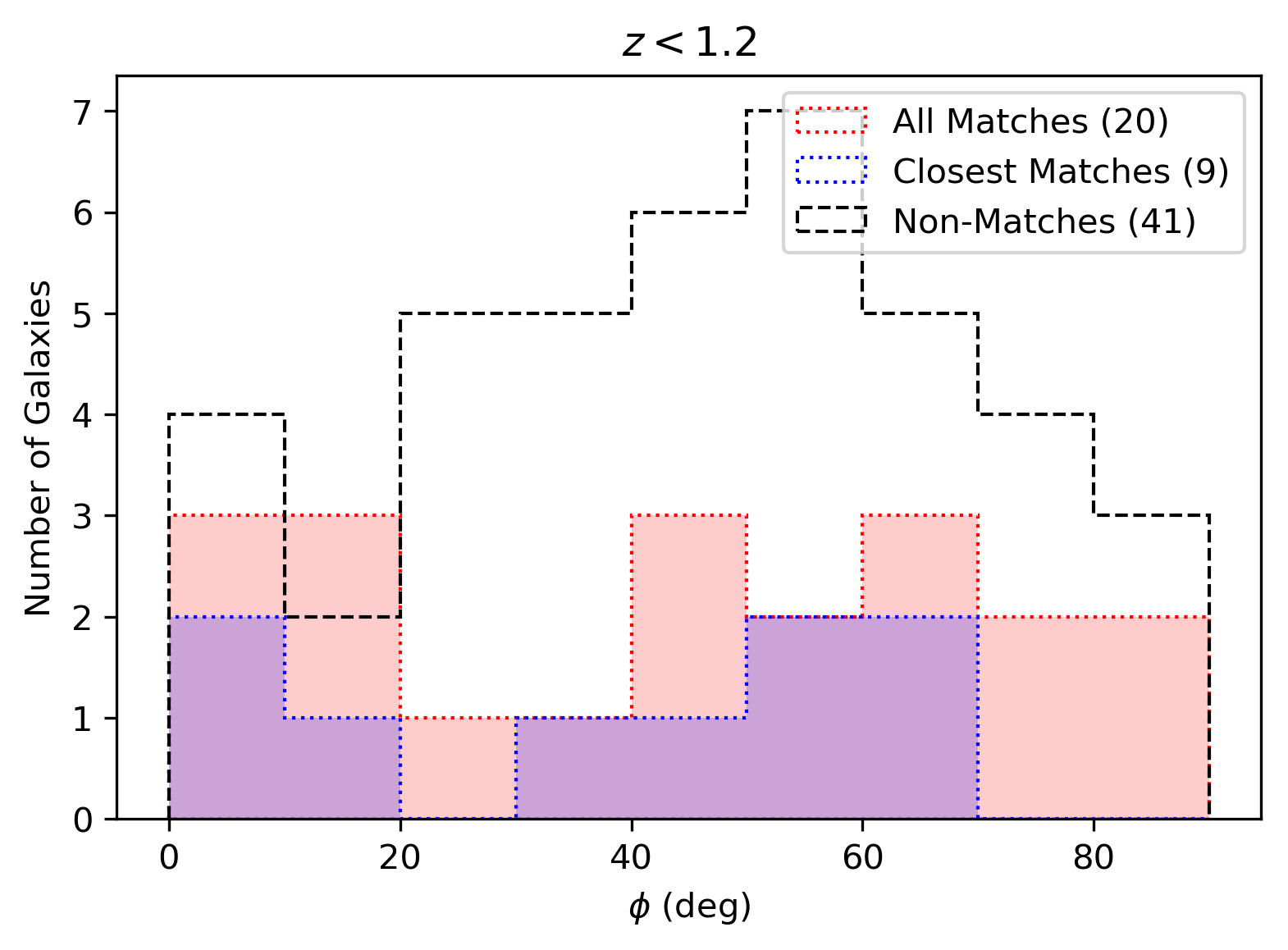}
    \includegraphics[width=0.45\textwidth]{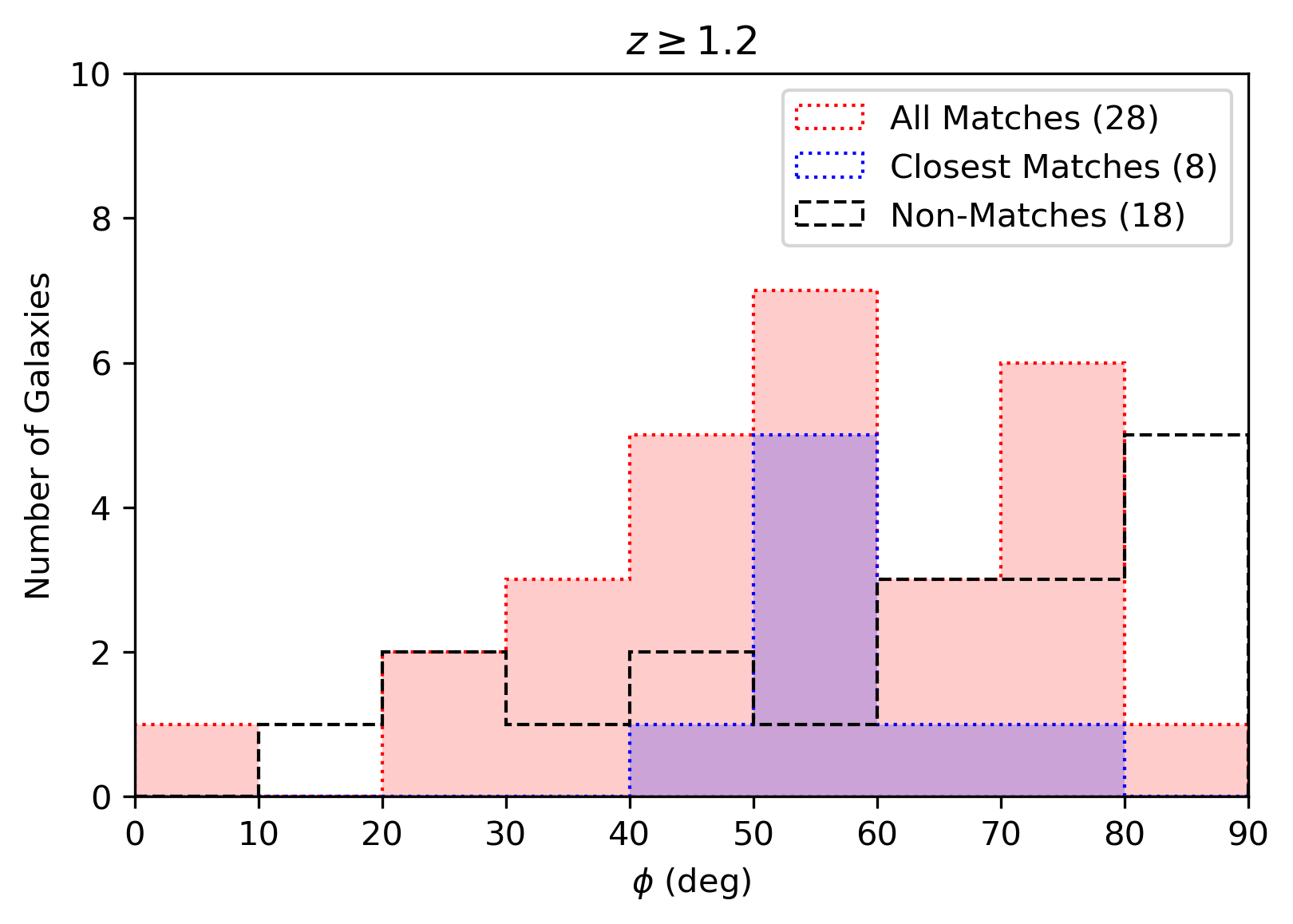}
    \caption{\small The dependence of the \ion{Mg}{2} azimuthal angle distribution on redshift. Lower redshift galaxies ($z<1.2$, $z_{median}=0.92$) are shown to the left, with higher redshift galaxies ($z\geq1.2$, $z_{median}=1.23$) at the right. Galaxies with  $\rho > 80$ kpc are excluded to reduce scatter, since the \ion{Mg}{2} distribution becomes more isotropic at larger impact parameters. The plot includes galaxies not matched with \ion{Mg}{2} absorption (black), closest matches (blue), and all matches (red).}
    \label{fig:ZAng}
\end{figure*}

Figure \ref{fig:AnzimuthalHist} presents the azimuthal angle distribution of the quasar sightlines with respect to foreground galaxies in this study with impact parameters less than 200 kpc. In cases where the multiple galaxies are spectroscopically matched to the same absorber, one might expect the galaxy with the smallest angular separation to the quasar to also have the strongest physical connection to the detected absorption. Thus, we have separately indicated our measurements for the nearest matches. The distribution of the closest matches is roughly bimodal, with peaks near $0 \degree$ and $60 \degree$. The fact that the combined azimuthal distribution for all of the galaxies in our sample exhibits an excess around $60\degree$ is presumably indicative of the absorption bias in our initial target selection combined with non-isotropic azimuthal distribution of \ion{Mg}{2} absorption in the CGM of galaxies. Galaxies with well-determined redshifts that are not matched to \ion{Mg}{2} (``non-matches") are found to exhibit an approximately flat distribution in azimuthal angle, as one would expect for a sample of galaxies with no preferred orientation relative to the background quasar.

For galaxies with more circular morphologies (axis ratios $b/a>0.6$), the azimuthal angle may not be well determined. We experimented with introducing a cut to remove galaxies with $b/a>0.6$ from our sample, but this did not affect the measured bimodality in a significant way.  This finding seems consistent with \citet{Bordoloi14a}, which reported that the wind geometry has negligible dependence on the inclination as compared to the dependence on azimuthal angle. 

The shape of galactic outflows caused by the combined winds of supernovae can be modelled as a cone whose axis is perpendicular to the plane of the galactic disk \citep{Schroetter15}. We find that the peak of the azimuthal distribution for the wind-pairs occurs at $60 \degree$, rather than $90 \degree$. In simulations performed by \citet{Nelson19} of wind-producing, low mass ($\log{(M_{*}/M_{\odot})} = 10.0$) galaxies, outflowing winds are initially launched with wide opening angles, but the gas becomes more collimated at impact parameters $\rho \sim 50$ kpc. In those simulations, these collimated winds form under-dense cavities along the minor axis, which are surrounded by a denser pressure front at an opening angle of $40 \degree - 50 \degree$. This would then produce stronger absorption when the quasar shines through the edge of the cone, consistent with the findings of \citet{Schroetter15} and \citet{Schroetter19}, and explaining our results as well.

We adopt similar nomenclature proposed by \citet{Schroetter15}, and refer to matches with $\phi>40 \degree$ as wind-pairs --- since winds produced during periods of high star formation are expected to flow out in this approximate angular region --- while matches at $<40 \degree$ are denoted as inflow-pairs, since gas inflowing from the intergalactic medium is likely to be accreted by the galaxy along the major axis. Of the 21 absorber-galaxy pairs in our sample with $\rho<80$~kpc, 81\% are categorized as "wind-pairs".  The probability of this observed distribution arising from a uniform angular distribution is less than 4\%.  The preferential enhancement of \ion{Mg}{2} that we find around the minor axes of galaxies in this sample is consistent with observations of late-type galaxies at $z\sim0.5$ \citep{Kacprzak12} and may be explained by gas aligned with the minor axis of galaxies having a higher metal enrichment, as would be expected in the case of star formation-driven outflows propagating perpendicular to the disks of galaxies.

In Figure \ref{fig:fracwind} we show, in bins of projected physical separation, the fraction of wind-pairs within the sample of \ion{Mg}{2} detections matched to foreground galaxies.  We find a significant excess of \ion{Mg}{2} detections with $\phi>40\degree$ at small impact parameters, which persists out to $\sim$80 kpc around foreground galaxies.  These observations are consistent with \citet{Schroetter19}, which recently reported the detection of a bimodal \ion{Mg}{2} distribution extending 60-80 kpc around star-forming galaxies at $z\sim1$.

As shown in Figure \ref{fig:AnzimuthalHist_BCut}, the bimodality in the azimuthal angle distribution of all galaxy-absorber matches (not just the closest matches) indeed strengthens when galaxies at impact parameters $\rho > 80$~kpc are excluded. This observation agrees with the findings of \citet{Bordoloi14a}, which reported that the distribution of \ion{Mg}{2} becomes increasingly isotropic at greater impact parameters, potentially indicating an increased contribution from in-falling material or satellite galaxies, or the maximum extent of cold gas in star formation-driven outflows.

Theoretical models predict that radiation pressure from galaxies with star formation rate surface densities of $\Sigma_{SFR}\gtrsim0.05$~M$_{\odot}$~yr$^{-1}$~kpc$^{-2}$ can drive outflows of $\sim10^{4}~K$ gas to distances of $50-100$~kpc around galaxies \citep{Murray2011, Hopkins2012}, and there are reasons to expect that the average morphologies of star formation-driven galaxy outflows may evolve over time. In an analysis of the IllustrisTNG simulation, \citet{Nelson19} claim that the collimation of outflows around star-forming galaxies emerges between the redshifts $z=2$ and $z=1$ and becomes more pronounced at lower redshifts. Simulations of outflows from $L_{*}$ galaxies in the \textit{FIRE} simulation \citep{Muratov15} also predict a weakening of the mass-loading factor of star formation-driven outflows with decreasing redshift, due in part to the deepening gravitational potential well in the center of the average galaxy halo and the lower frequency of gas-rich mergers triggering episodic bursts of star formation. Thus we might expect to observe an evolution in both the collimation and extent of outflows around star-forming galaxies in the range $0<z<2$, with outflows extending farther into the halos of galaxies at $z>1$, compared to at lower redshifts.  

\begin{figure*}
\centering
    \includegraphics[width=0.45\textwidth]{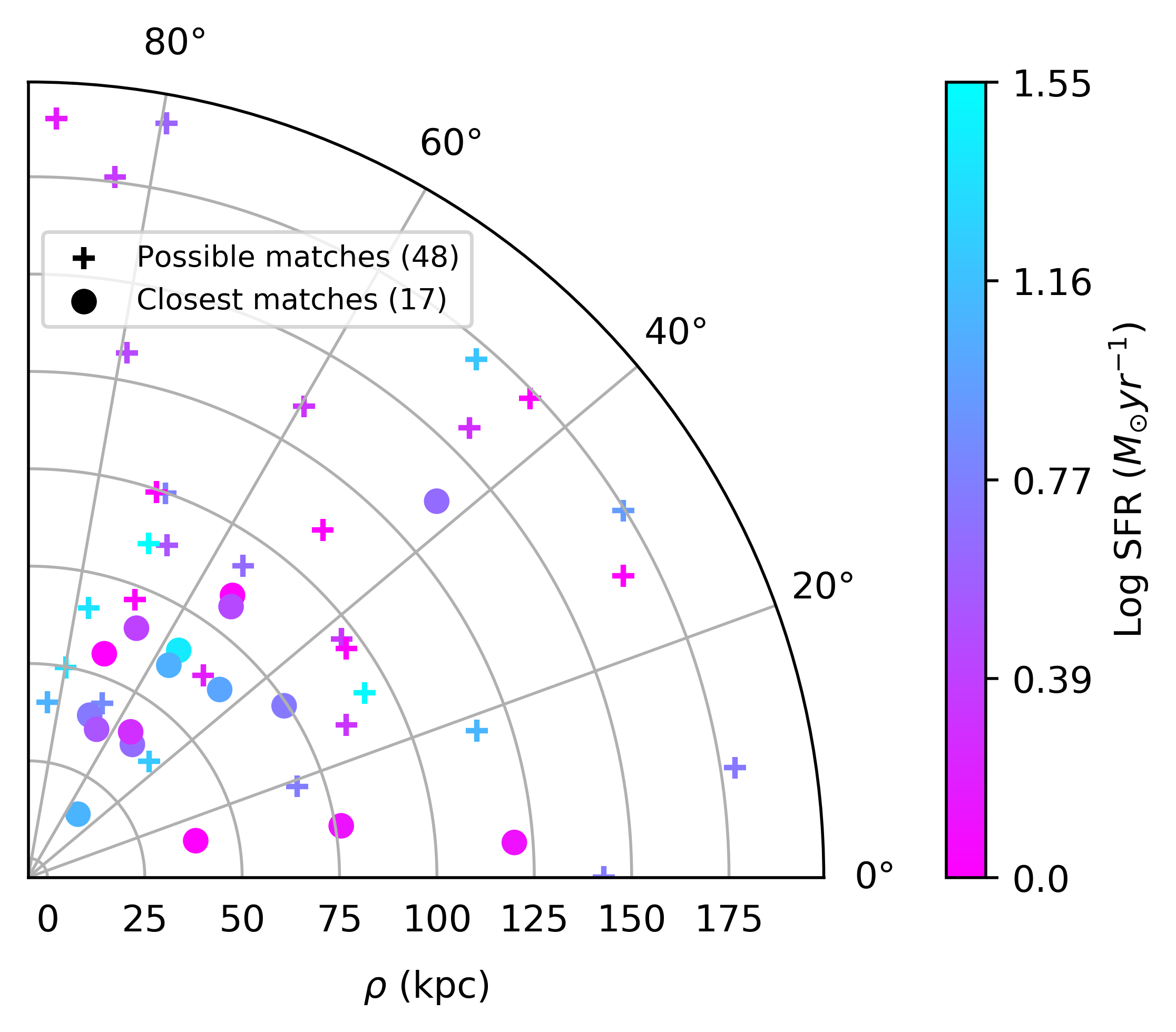}
    \includegraphics[width=0.45\textwidth]{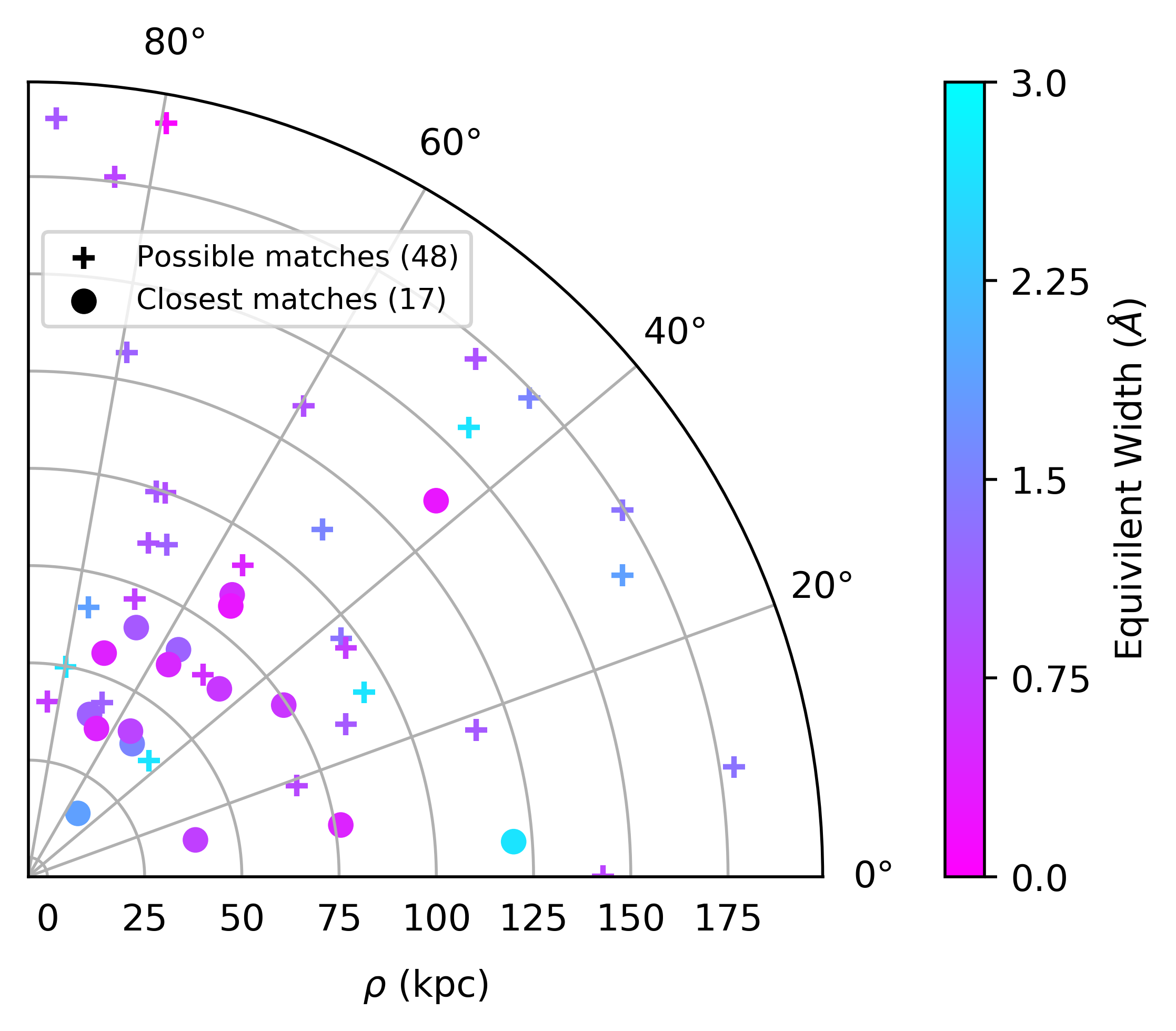}
    \caption{\small The observed relation between azimuthal angle and impact parameter for galaxies matched to \ion{Mg}{2}, shown with color denoting the star formation rate (left) and the rest-frame equivalent width of the 2796\AA~\ion{Mg}{2} absorption line (right). 
    }
    \label{fig:DistAng}
\end{figure*}

\citet{Kacprzak11b} and \citet{Bordoloi14a} report that the bimodality of the azimuthal distribution disappears at $\rho > 40$~kpc at $z\sim0.5$, while our study --- at significantly higher redshift ($z \sim 1.2$) --- finds persistent bimodality out to $\rho=80$~kpc.   Combined with the similar findings at $z\sim1$ from \citet{Schroetter19}, these observational results appear to support theoretical predictions of star formation-driven outflows reaching larger physical distances around galaxies at earlier times.

\begin{figure}[t!]
    \includegraphics[width=1.0\columnwidth]{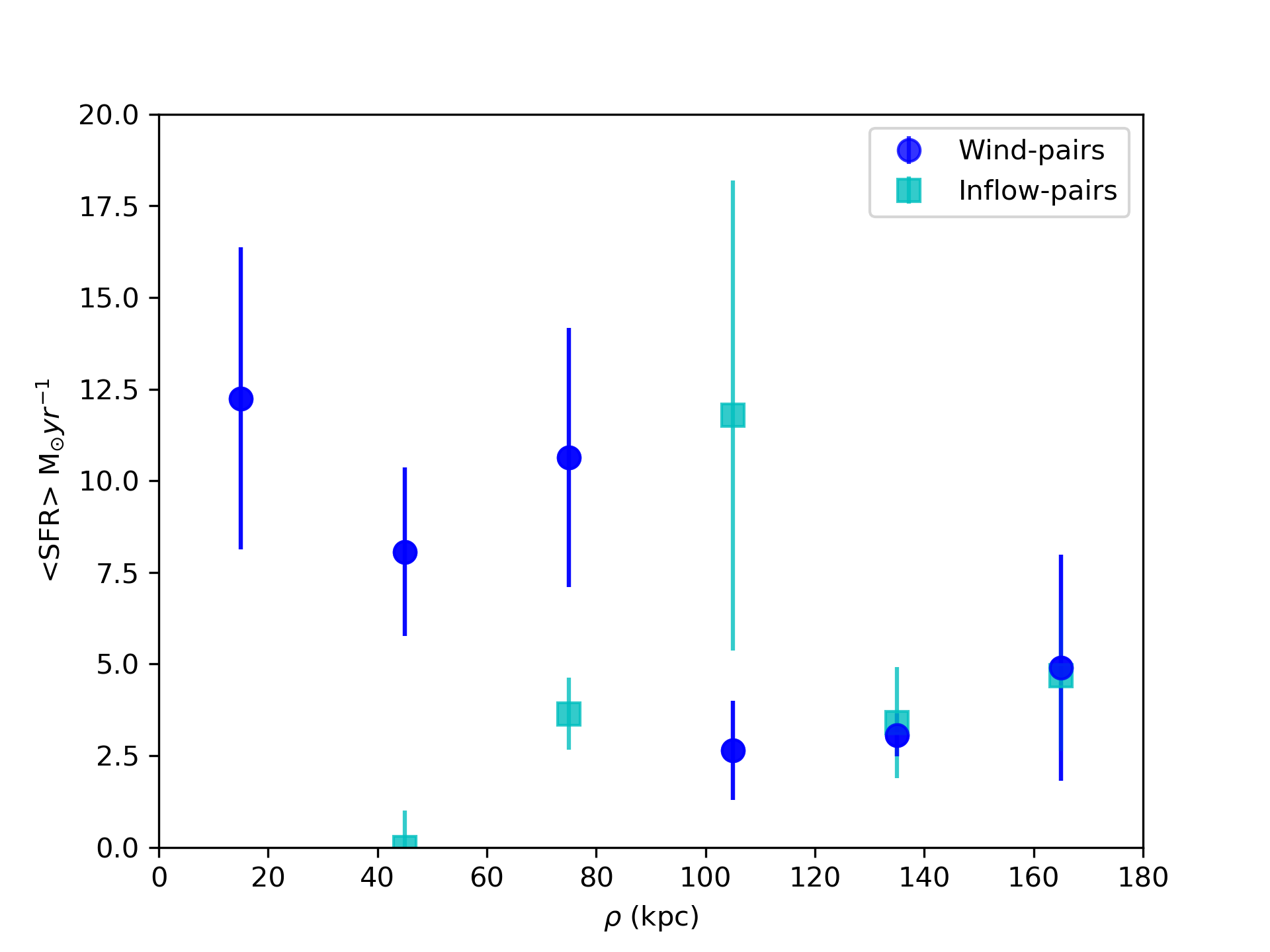}
    \caption{\small The mean SFR of galaxies matched to \ion{Mg}{2} absorption, shown as a function of impact parameter ($\rho$). The data are subdivided by azimuthal angle to wind-pairs ($\phi\geq40\degree$) and inflow-pairs ($\phi<40\degree$). Wind-pairs with $\rho\lesssim80$ kpc have significantly higher mean galaxy SFRs compared to inflow-pairs at similar impact parameters and wind-pairs at $\rho>100$ kpc.}
    \label{fig:SFR_angdep}
\end{figure}

In Figure~\ref{fig:ZAng} we split our galaxy-absorber pairs at the median redshift ($z=1.2$) into high and low redshift sub-samples with median redshifts of z=0.92 and z=1.23, respectively. Plotting the azimuthal distribution of each sub-sample indicates that the fraction of \ion{Mg}{2} absorbers detected with $\phi>40\degree$ is marginally greater at higher redshift.  Amongst all of the possible matched galaxy-absorber pairs in our sample 60\%  (12/20) can be classified as wind-pairs at $z<1.2$, compared to 79\% (22/28) at $z\geq1.2$. Despite these reasonably small size of these samples, which limits our ability to draw conclusions about the possible redshift evolution in the azimuthal distribution (Figure ~\ref{fig:ZAng}), the large fraction of wind-pairs in the higher redshift sub-sample is significant. A K-S test indicates that the azimuthal distribution of the $z\geq1.2$ sub-sample, which peaks at 55\degree, has a $<3$\% likelihood of being drawn from a random distribution.

In Figure~\ref{fig:DistAng} we present the distribution of our galaxy-absorber pairs as a function of impact parameter and azimuthal angle, color-coded separately by \ion{Mg}{2} equivalent width ($W_{r}$) and galaxy SFR. The majority (79\%) of galaxies with SFRs above the mean of this sample have \ion{Mg}{2} detections at azimuthal angles greater than 40\degree, indicating an outflow origin. Interestingly, in contrast to many other lines of evidence linking higher equivalent width \ion{Mg}{2} absorbers to star formation-driven outflows, we find no significant correlation between $W_{r}$ and azimuthal angle in our sample. 

Figure~\ref{fig:SFR_angdep} presents the average SFR of galaxies matched to \ion{Mg}{2} absorption in bins of impact parameter, separately for wind-pairs and inflow-pairs.  Wind-pairs with $\rho\lesssim80$ kpc have significantly higher mean galaxy SFRs compared to wind-pairs at $\rho>100$ kpc. While the small number of inflow-pairs restricts what conclusions can be drawn about their global properties, these data appear to suggest that they may also be associated with  lower star formation rates compared to wind-pairs. A larger sample of inflow pairs would be needed to confirm this possible trend.

Interestingly, despite the bimodality of the azimuthal distribution at at $\rho<80$ kpc, which is suggestive of inflows and outflows, a significant fraction of the detected wind-pairs have low star formation rates. This result could potentially be explained by the sporadic nature of star formation activity; the period of rapid star formation that generated strong winds could have quieted by the time the outflowing gas reaches the quasar sightline. It is also possible that some of these absorbers may have been assigned to galaxies with matching redshifts that are not their primary hosts.

\subsection{Modeling galaxy contributions to W$_{r}$} \label{sec:modeling}

Several factors -- including the origins of the gas and the properties of its host galaxy -- are understood to contribute to the equivalent width of the \ion{Mg}{2} detected in the circumgalactic medium of galaxies. Determining the relative effects of the many contributing factors from the equivalent width measurement of a single galaxy-absorber pair may be impossible with one individual detection. However, with a statistical sample of galaxy-absorber pairs for which those contributions are measured, we can develop a model that describes the average expected contribution of each factor to the measured equivalent width. 

As previously discussed in Section \ref{sec:EW_b}, the anti-correlation between $W_r$ and impact parameter is well established, but there is considerable scatter around the best-fit trend line (Figure \ref{fig:WvsRho}). Various properties of the host galaxy (e.g., star formation rate, azimuthal angle, stellar mass) and its environment have been shown to correlate with $W_r$ \citep{Chen10b,Kacprzak11a,B12b, Nielsen2013,Churchill13}, and are therefore expected to contribute to the scatter in the $W_{r}-\rho$ relation.

Our observations provide additional information about the galaxies matched with absorption -- namely azimuthal angle and environment -- that may allow us to better predict the equivalent width of a galaxy-absorber pair. Using these measurements, we have created a model that adopts the previously established relationship between $W_r$ and impact parameter at high redshifts (discussed in section \ref{sec:EW_b}) while also folding in the observed correlation between $W_r$ and azimuthal angle. Incorporating SFR information did not significantly improve the predictions of equivalent width, so we ultimately excluded it from our model. This lack of importance of galaxy SFR may be surprising, because past studies at intermediate redshift \citep[e.g.,][]{Zibetti07, Menard08, NSM10, Menard11, Bordoloi11, Menard12, Bordoloi14b, Lan2018} have found a strong correlation between $W_r$ and the star formation rates of matching galaxies. There are a few possible explanations for this result. Since most $z \sim 1$ galaxies have relatively high star formation rates, we could be under-sampling galaxies with low SFRs. In that case, we may be lacking the dynamic range required to accurately model the correlation of SFR with $W_r$. This problem may be further compounded by the lack of dust correction in our SFR estimates, as previously discussed. The small size of our data set could also obscure the contributions of SFR to $W_r$.  We note that while stellar mass is also expected to affect the covering fraction and distribution of \ion{Mg}{2} absorption, we are currently unable to account for this additional affect.

\begin{figure}[t!]
    \includegraphics[width=1.0\columnwidth]{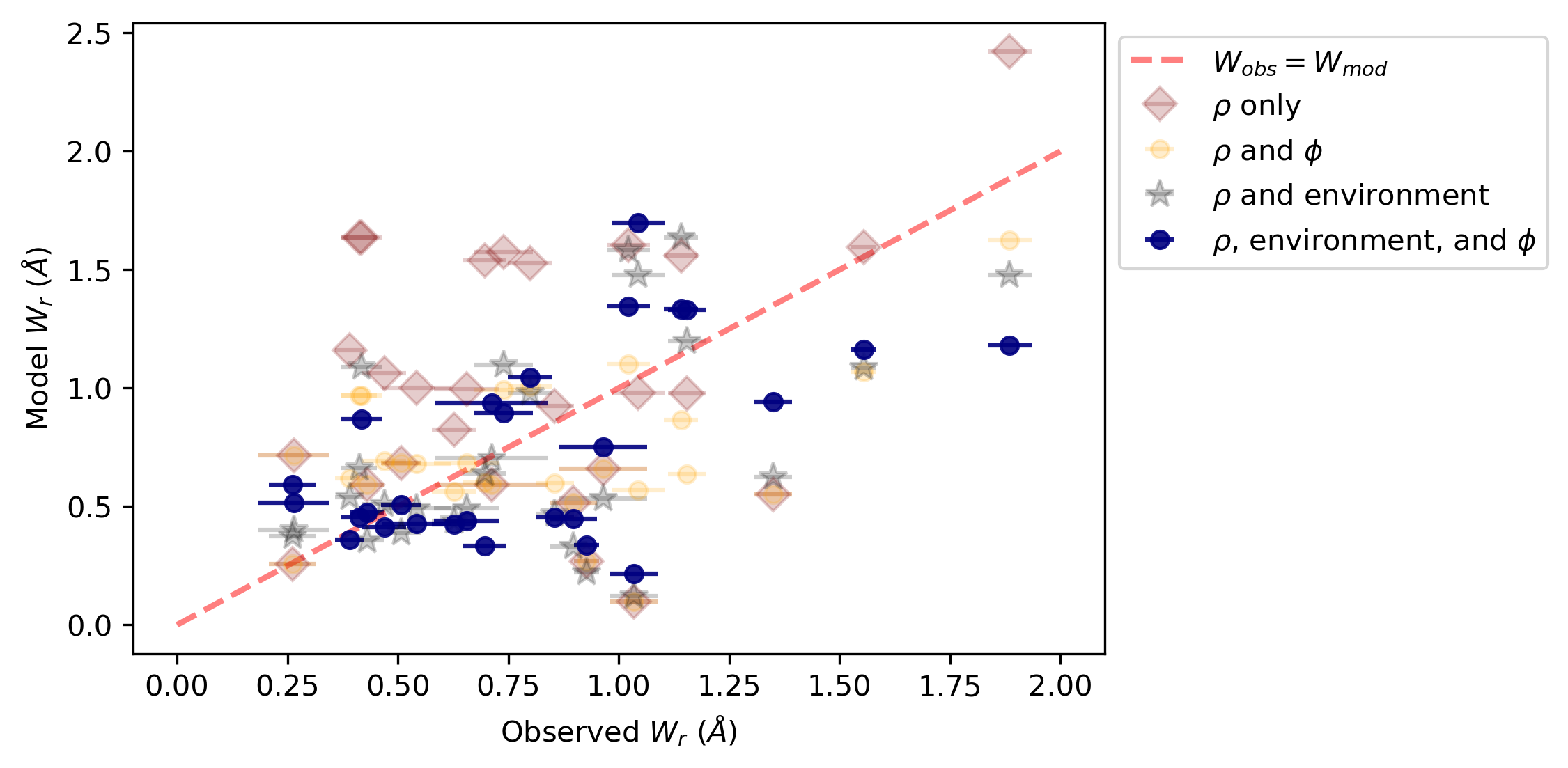}
    \caption{The predicted versus observed $W_r$ of each absorber using three different models. Red shows the model that only takes impact parameter into account (as discussed in section \ref{sec:EW_b}) ($\chi^{2}=16.5$), orange takes angle and impact parameter into account ($\chi^{2}=5.66$), black includes the effects of impact parameter and environment ($\chi^{2}=5.17$), while blue includes the effects of all three parameters ($\chi^{2}= 4.23$).} 
    \label{fig:model_progression}
\end{figure}

Figure~\ref{fig:model_progression} shows the results of four different models of $W_r$: one that only includes impact parameter (section 4.2), one that includes impact parameter and environmental information, one that takes impact parameter and azimuthal angle into account, and a final, three-parameter model (impact parameter, environment, and azimuthal angle). We find that our ability to predict the observed $W_r$ is strengthened when both azimuthal angle and environmental information are included. This is consistent with findings by \citet{Kacprzak12}, which reported that azimuthal angle is an important factor in determining a galaxy's contribution to $W_r$.

\begin{figure}[t!]
\includegraphics[width=1.0\columnwidth]{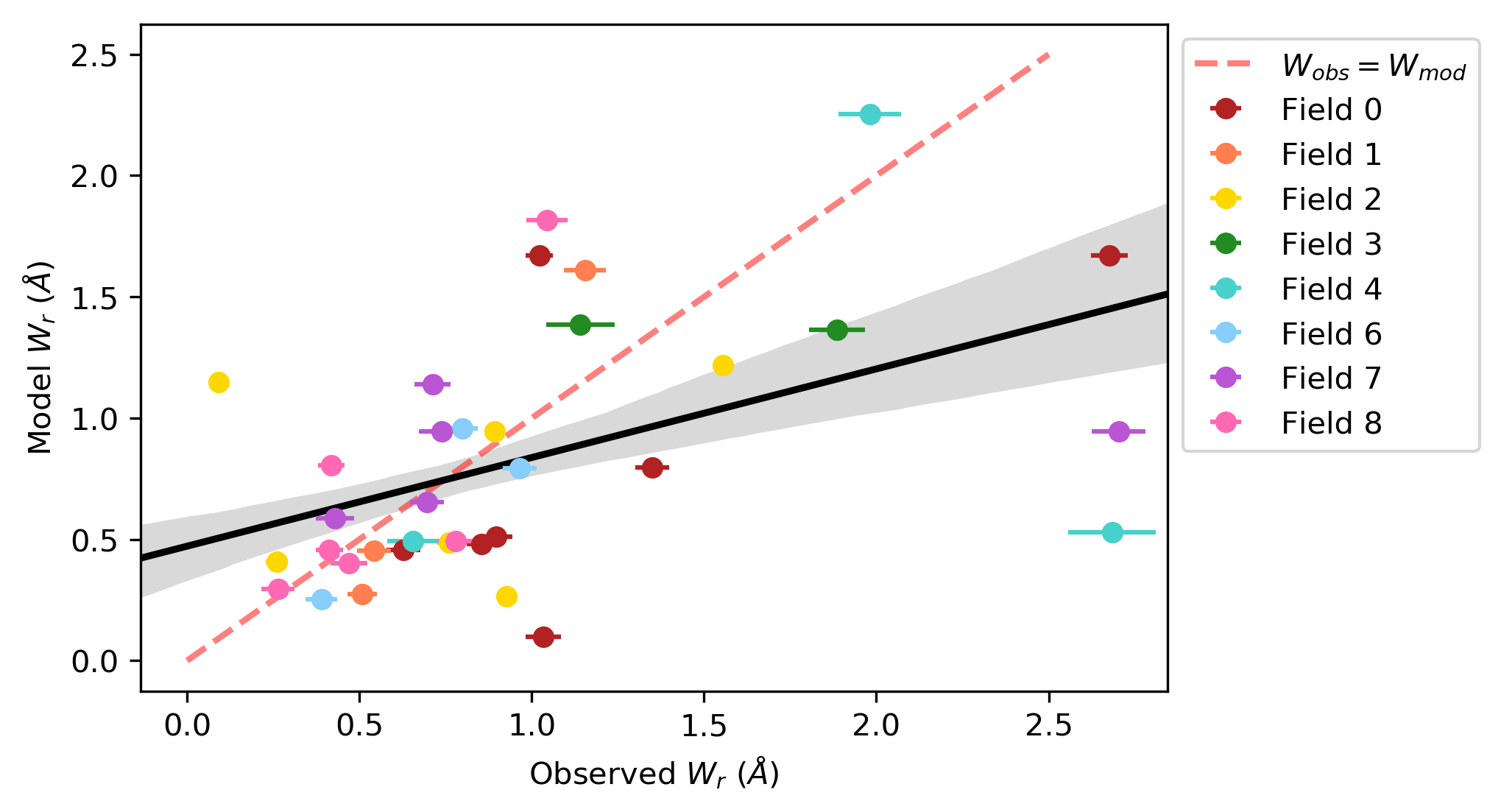}
\includegraphics[width=1.0\columnwidth]{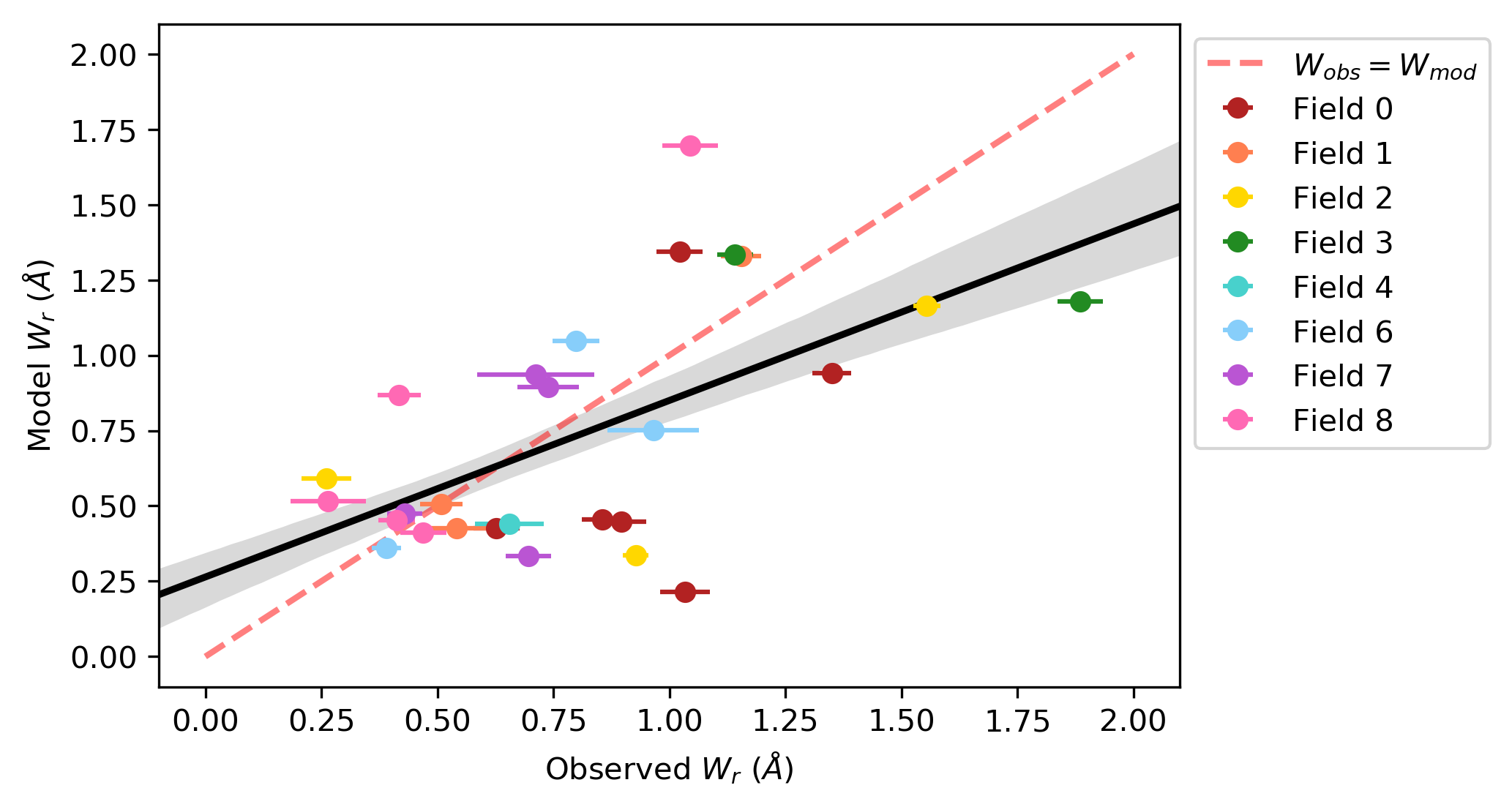}
\caption{\small The predicted versus observed $W_r$ of each absorber. The predicted $W_r$ is acquired using Equations \ref{eq:eqn_model} and \ref{eq:eqn_model_highrho}, which sum over all galaxies matching in redshift to a \ion{Mg}{2} detection. The top panel shows all absorbers, while the bottom panel excludes ultra-strong absorbers, defined as $W_r>2.0$\AA~. The trend line for the model is plotted in black with a shaded 1-$\sigma$ region, and a one-to-one relation is given by the dashed red line. The $\chi^{2}/(n-1)$ for this model is $0.325$ (top) and $0.163$ (bottom).}.\label{fig:Model_RandAngles}
\end{figure}

In order to match the bimodality observed in the azimuthal distribution (Figure \ref{fig:AnzimuthalHist}), we modeled the correlation between azimuthal angle and $W_r$ as two Gaussian functions with peaks at $60 \degree$ and $0 \degree$. For galaxies at impact parameters $\rho < 80$ kpc, we model the summed contributions to the \ion{Mg}{2} equivalent width as follows:

\begin{equation}
W_r(\rho<80~\text{kpc}) = \sum\limits_{k=1}^n 10^{a * \rho + b} \bigg[ \alpha_1 e^{ -\frac{(\phi-60)^2}{2(s_1^2)}} + \alpha_2 e^{-\frac{\phi^2}{2(s_2^2)}} \bigg], 
\label{eq:eqn_model}
\end{equation}

\noindent where $\phi$ is the azimuthal angle of the galaxy, $\rho$ is its impact parameter, $n$ is the number of galaxies matched to absorption in the quasar sightline with $\rho<80$~kpc, $a = -0.003$ and $b = 0.0$, $\alpha_1 = 0.43$ and $\alpha_2 = 0.56$, and the standard deviations of the Gaussian functions are $s_1=44.5$ and $s_2=43.4$. 

Since the distribution of \ion{Mg}{2} becomes isotropic at larger impact parameters ($\rho > 80$ kpc), as discussed in Section \ref{sec:azang}, we model the contribution of galaxies at $\rho \geq 80$ kpc without azimuthal considerations, as follows:

\begin{equation}
W_r(\rho\geq80~\text{kpc}) = \sum\limits_{k=1}^n 10^{a * \rho + b}. 
\label{eq:eqn_model_highrho}
\end{equation}

So the total modeled $W_r$ is the summative contribution of all galaxies with impact parameters $\rho<80$ kpc (equation \ref{eq:eqn_model}) and $\rho>80$ kpc (equation \ref{eq:eqn_model_highrho}):

\begin{equation}
W_{r\text{ }total} = W_r(\rho<80~\text{kpc}) + W_r(\rho\geq80~ \text{kpc})
\label{eq:eqn_model_total}
\end{equation}

This model sums the $W_r$ contributions of all galaxies that are spectroscopically matched to a given absorption feature. It uses the impact parameter to predict the maximum $W_r$ (in \AA) contribution of a given galaxy, then uses the azimuthal angle to modify that result.  $a$, $b$, $\alpha_1$, $\alpha_2$, $s_1$, and $s_2$ were treated as free parameters in the model and are fit for our data when ultra-strong absorbers ($W_r > 2$\AA) are excluded. 

As shown in Figure~\ref{fig:Model_RandAngles}, the model performs worse when ultra-strong absorbers are included. The strongest absorbers are the most likely to have components at very small impact parameters, which may be unresolved from the quasar PSF. Such cases would result in some galaxy-absorber pairs being excluded from the input to the model and could lead to a worse prediction of W$_{r}$. Galaxies with uncertain redshifts (Table \ref{tbl-absorbers}), which we have excluded, could also be contributing to the overall scatter about the best-fit parameterization, which worsens for absorbers with $W_r > 2$\AA. The \ion{Mg}{2} absorbers modeled in Figure~\ref{fig:Model_RandAngles} are color-coded by the field in which they were detected.

What sets this model apart from previous attempts to understand the relationship between $W_r$ and impact parameter is the inclusion of both azimuthal angles and environmental information through the summed contributions of all galaxies matched to absorption. We found that only considering the closest matching galaxy for each absorber, rather than all of the matching pairs, worsened the agreement between the model prediction and the detected $W_r$ (Figure~\ref{fig:model_progression}). There is potential for this model to be further refined in the future by fitting to a larger sample and by including more precisely determined parameters, such as the dust-corrected star formation rate and stellar mass.

\section{Summary}

We have presented first results from an 18 orbit HST program that obtained WFC3/IR G141 grism and F140W direct imaging observations of galaxies around the nine most \ion{Mg}{2}-rich quasar sightlines in the SDSS. These data have enabled a study of the morphologies, azimuthal angles, star formation rates, star formation rate surface densities, and environments in a large sample of typical \ion{Mg}{2} absorbing galaxies at $0.64<z<1.6$. We were able to measure morphologies for 107 galaxies and star formation rates for 98 galaxies. Only one field was fully excluded from our analysis due to the crowding of several galaxies with projected separations of $<5$\arcsec~from the quasar, which caused catastrophic contamination in the G141 grism observations.

We report an exceptionally high rate of detection for candidate \ion{Mg}{2} host galaxies; 89\% (34/38) of the targeted \ion{Mg}{2} absorption systems were confidently matched to at least one galaxy with $|\Delta z|/(1+z_{MgII})<0.006$ within 200 kpc. These observations have helped to expand the limited sample of spectroscopically confirmed \ion{Mg}{2}-galaxy pairs at $z>1$, which, together with compilations of observations in the literature at lower redshift, indicate a significant redshift evolution in the distribution of \ion{Mg}{2} around absorption-selected galaxies.

Nearly half of the absorbers in our targeted sample matched in redshift to two or more galaxies, and the mean \ion{Mg}{2} rest-frame equivalent width ($W_{r}$) of absorbers matched with groups is  greater than that of absorbers matched to isolated galaxies. The sample of galaxies matched to \ion{Mg}{2} absorption were found to have a significantly higher mean star formation rate, and a marginally higher mean $\Sigma_{\rm{SFR}}$, compared to galaxies that were not matched to \ion{Mg}{2} absorption.  

We detect a bimodal azimuthal angle distribution of \ion{Mg}{2} around galaxies in our sample.  Most of the galaxy-absorber pairs were detected within 50\degree~of the minor axis, suggestive of an origin in star formation-driven outflows. The bimodality of the \ion{Mg}{2} azimuthal angle distribution extends to impact parameters of ($\rho \sim 80$~kpc), in agreement with other findings of extended \ion{Mg}{2} around galaxies at $z\sim1$ \citep{Schroetter19} and supporting predictions from recent IllustrisTNG simulations \citep{Nelson19}. We also find that the signature of wind-driven outflows in the azimuthal angle distribution is more prominent in the higher-redshift half of our sample.

Finally, we present a new model that uses impact parameter, azimuthal angle, and galaxy environment to model the detected W$_{r}$ of \ion{Mg}{2}. The results of our modeling indicates that accounting for azimuthal angle of the absorption around galaxies and the cumulative effect of multiple galaxies in a group environment results in a better understanding of the detected \ion{Mg}{2} equivalent width.

In the future, deeper multi-wavelength broadband imaging of these high value fields would enable stellar population synthesis modeling, enabling the further examination of the geometry of \ion{Mg}{2} absorption as a function of galaxy age, mass, and virial radius. Such observations would also provide photometric redshift constraints for the galaxies in this sample that lack strong emission lines, facilitating improved certainty in absorber pairing and group membership determinations.  Higher resolution spectroscopy of both the galaxies and quasar absorption lines would also be useful for improving the matching of galaxies to absorption line components in group environments.

\section*{Acknowledgments}

We would like to thank the anonymous referee for their detailed feedback, and Dylan Nelson for helpful discussions. This work would not have been possible without the publicly available data from the Sloan Digital Sky Survey and the GRIZLI WFC3/IR Grism reduction software\footnote{https://github.com/gbrammer/grizli}.  Our observations and analysis were supported by the HST G0-13482 program with the NASA/ESA Hubble Space Telescope, which is operated by the Association of Universities for Research in Astronomy, Inc., under NASA contract NAS5-26555. This research has made extensive use of NASA's Astrophysics Data System Bibliographic Services and of open source scientific Python libraries, including PyFITS and PyRAF produced by the Space Telescope Science Institute, which is operated by AURA for NASA.  

Funding for the Sloan Digital Sky Survey IV has been provided by the Alfred P. Sloan Foundation, the U.S. Department of Energy Office of Science, and the Participating Institutions. SDSS-IV acknowledges
support and resources from the Center for High-Performance Computing at
the University of Utah. The SDSS web site is www.sdss.org.

SDSS-IV is managed by the Astrophysical Research Consortium for the 
Participating Institutions of the SDSS Collaboration including the 
Brazilian Participation Group, the Carnegie Institution for Science, 
Carnegie Mellon University, the Chilean Participation Group, the French Participation Group, Harvard-Smithsonian Center for Astrophysics, 
Instituto de Astrof\'isica de Canarias, The Johns Hopkins University, Kavli Institute for the Physics and Mathematics of the Universe (IPMU) / 
University of Tokyo, the Korean Participation Group, Lawrence Berkeley National Laboratory, 
Leibniz Institut f\"ur Astrophysik Potsdam (AIP),  
Max-Planck-Institut f\"ur Astronomie (MPIA Heidelberg), 
Max-Planck-Institut f\"ur Astrophysik (MPA Garching), 
Max-Planck-Institut f\"ur Extraterrestrische Physik (MPE), 
National Astronomical Observatories of China, New Mexico State University, 
New York University, University of Notre Dame, 
Observat\'ario Nacional / MCTI, The Ohio State University, 
Pennsylvania State University, Shanghai Astronomical Observatory, 
United Kingdom Participation Group,
Universidad Nacional Aut\'onoma de M\'exico, University of Arizona, 
University of Colorado Boulder, University of Oxford, University of Portsmouth, 
University of Utah, University of Virginia, University of Washington, University of Wisconsin, 
Vanderbilt University, and Yale University.

\clearpage

\end{document}